# Widespread Rapid Quenching at Cosmic Noon Revealed by JWST Deep Spectroscopy

Minjung Park[1], Sirio Belli[2], Charlie Conroy[1], Benjamin D. Johnson[1], Rebecca L. Davies[3,4], Joel Leja[5,6,7], Sandro Tacchella[8,9], J. Trevor Mendel[4,10], Chloë Benton[11], Letizia Bugiani[2], Razieh Emami[1], Amir H. Khoram[2], Yijia Li[5,6], Gabriel Maheson[8,9], Elijah P. Mathews[5,6,7], Rohan P. Naidu[12], Erica J. Nelson[11], Bryan A. Terrazas[13], and Rainer Weinberger[14]

[1] Center for Astrophysics | Harvard & Smithsonian, Cambridge, MA 02138, USA
[2] Dipartimento di Fisica e Astronomia, Università di Bologna, Bologna 40129, Italy
[3] Centre for Astrophysics and Supercomputing, Swinburne University of Technology, Hawthorn, Victoria 3122, Australia
[4] ARC Centre of Excellence for All Sky Astrophysics in 3 Dimensions (ASTRO 3D), Australia
[5] Department of Astronomy & Astrophysics, The Pennsylvania State University, University Park, PA 16802, USA
[6] Institute for Gravitation and the Cosmos, The Pennsylvania State University, University Park, PA 16802, USA
[7] Institute for Computational & Data Sciences, The Pennsylvania State University, University Park, PA 16802, USA
[8] Kavli Institute for Cosmology, University of Cambridge, Cambridge CB3 0EZ, UK
[9] Cavendish Laboratory, University of Cambridge, Cambridge CB3 0EZ, UK
[10] Research School of Astronomy and Astrophysics, Australian National University, Canberra 2611, ACT, Australia
[11] Department for Astrophysical and Planetary Science, University of Colorado, Boulder, CO 80305, USA
[12] MIT Kavli Institute for Astrophysics and Space Research, Cambridge, MA 02139, USA
[13] Department of Physics & Astronomy, Oberlin College, Oberlin, OH 44074, USA
[14] Leibniz Institute for Astrophysics, An der Starnwarte 16, 14482 Potsdam, Germany
*Received 2024 April 27; revised 2024 September 19; accepted 2024 September 20; published 2024 November 13*

## Abstract

Massive quiescent galaxies in the young Universe are expected to be quenched rapidly, but it is unclear whether they all experience starbursts before quenching and what physical mechanism drives rapid quenching. We study 14 massive quiescent galaxies ($\log(M_\star/M_\odot) > 10$) at $z \sim 2$ selected from a representative sample of the Blue Jay survey. We reconstruct their star formation histories (SFHs) by fitting spectral energy distribution models to the JWST/NIRSpec $R \sim 1000$ spectra. We find that massive quiescent galaxies can be split into three categories with roughly equal numbers of galaxies according to their SFHs: (1) relatively old galaxies quenched at early epochs; (2) galaxies that are rapidly and recently quenched after a flat or bursty formation history (depending on the assumed prior); and (3) galaxies that are rapidly and recently quenched after a major starburst. Most recently quenched galaxies show neutral gas outflows, probed by blueshifted Na I D absorption, and ionized gas emission, with line ratios consistent with active galactic nucleus (AGN) diagnostics. This suggests that AGN activity drives multiphase gas outflows, leading to rapid quenching. By tracing back the SFHs of the entire sample, we predict the number density of massive quiescent galaxies at $z = 4$–6: $n = (1.5$–$6.0) \times 10^{-5}\,\mathrm{Mpc}^{-3}$. The two old massive quiescent galaxies in our sample appear to have extremely early formation and quenching ($z \gtrsim 6$) and are possibly descendants of early post-starbursts at $z > 3$. These galaxies still show neutral gas reservoirs and weak H$\alpha$ emission, perhaps because the ejective AGN feedback that caused rapid quenching has weakened over time.

*Unified Astronomy Thesaurus concepts:* Galaxies (573); Galaxy formation (595); Galaxy evolution (594); Galaxy quenching (2040)

## 1. Introduction

Star-forming activity is a key characteristic of galaxies that governs their growth. In the local Universe, it is also tightly related to other galaxy properties, such as stellar mass, size, and morphology (e.g., I. K. Baldry et al. 2004; S. Wuyts et al. 2011; A. F. L. Bluck et al. 2014). One of the important questions in galaxy formation is understanding why and how galaxies stop forming stars and become quiescent. Several quenching mechanisms have been proposed over decades (see A. Man & S. Belli 2018 and references therein), which can be broadly classified into two main processes with different timescales: "rapid" (a timescale on the order of a few hundred Myr) and "slow" (>1 Gyr timescale) quenching (e.g., P.-F. Wu et al. 2018; S. Belli et al. 2019; V. Wild et al. 2020). Typically, the mechanisms involving the gas removal process—for example, ram pressure stripping (e.g., J. E. Gunn & J.R.I. Gott 1972) or ejective active galactic nucleus (AGN) feedback (e.g., J. Silk & M. J. Rees 1998; T. Di Matteo et al. 2005)—are thought to quench galaxies on short timescales. On the other hand, gas-heating processes—such as virial shock heating (e.g., S. D. M. White & M. J. Rees 1978; D. Kere et al. 2005) or thermal AGN feedback (e.g., D. J. Croton et al. 2006; R. S. Somerville et al. 2008)—are thought to take much longer to quench galaxies. There have been many studies constraining the quenching timescales. At lower redshifts ($z < 1$), galaxies seem to have a broad range of quenching timescales, from several hundred Myr to a few Gyr (e.g., Y. Peng et al. 2015; J. Trussler et al. 2020; S. Tacchella et al. 2022a), suggesting multiple quenching mechanisms behind them.

The "post-starburst" galaxies (PSBs) are thought to be in a rapid transition from star-forming to quiescence. In the local Universe, they are a rare population comprising only a few percent of the quiescent population (e.g., V. Wild et al. 2016; P.-F. Wu et al. 2018). They are first introduced as E+A or k+A galaxies that are dominated by young stars but not currently forming stars (e.g., A. Dressler & J. E. Gunn 1983). The most conventional way to select these populations is using Balmer absorption, indicating young stars but requiring weak emission (e.g., H$\alpha$ or [O II] emission lines) to select against ongoing star formation (e.g., A. I. Zabludoff et al. 1996; A. Dressler et al. 1999; T. Goto 2005; K. D. French et al. 2015). At higher redshifts, many studies have found age trends in the rest-frame $UVJ$ diagram and selected young quiescent galaxies based on their locations in the $UVJ$ plane (e.g., K. E. Whitaker et al. 2012a; S. Belli et al. 2019; K. A. Suess et al. 2020) or using other photometric selections (e.g., V. Wild et al. 2016, 2020). However, it has been debated whether these young quiescent galaxies are truly PSBs or just rapidly quenched galaxies (RQs), which can only be revealed by reconstructing their detailed star formation histories (SFHs).

At high redshifts ($z > 3$), PSBs are expected to be more common. Because the Universe is only a few Gyr old, rapid quenching is "required" for quiescent galaxies. Also, massive quiescent galaxies are likely to have experienced intense starbursts to build up their high stellar masses. Mergers are more frequent and galaxies tend to be gas-rich at high redshifts, providing a good condition to induce starbursts (e.g., J. C. Mihos & L. Hernquist 1996). Indeed, many studies have found that the fraction of young quiescent galaxies increases with redshift (e.g., K. E. Whitaker et al. 2012a; V. Wild et al. 2016; S. Belli et al. 2019; D. J. Setton et al. 2023). Several studies have confirmed that almost all quiescent galaxies observed at $z > 3$ appear to have the PSB nature (e.g., C. D'Eugenio et al. 2020; K. D. French 2021; T. J. Looser et al. 2023). However, it is not entirely clear exactly what fraction of quiescent galaxies at the intermediate redshift $z \sim 2$ are rapidly or slowly quenched and how many of the RQs are PSBs.

The ejective AGN feedback (e.g., J. Silk & M. J. Rees 1998; T. Di Matteo et al. 2005) is thought to be an important mechanism explaining the rapid quenching of high-$z$ massive quiescent galaxies (e.g., R. Weinberger et al. 2018); supermassive black holes (SMBHs) drive powerful galactic outflows that blow away gas, which eventually leads to quenching. Galactic outflows, probed by ionized gas emission, have been found almost ubiquitously in massive star-forming galaxies at $z > 1$ (e.g., N. M. Förster Schreiber et al. 2019; G. C. K. Leung et al. 2019; R. L. Davies et al. 2020). However, it has been pointed out that the ejected mass in the ionized phase is insufficient to fully suppress star formation (e.g., I. Lamperti et al. 2021). A larger amount of ejected gas is thought to be in neutral and molecular phases, but observing outflows in these phases has been challenging at high redshifts (e.g., S. Veilleux et al. 2020). With the unprecedented sensitivity of JWST, the Na I D $\lambda$5891, $\lambda$5897 absorption has been detected in many high-$z$ galaxies, which can be used as a probe of neutral gas. Recent studies of S. Belli et al. (2024) and F. D'Eugenio et al. (2024) have detected neutral gas outflows in two PSBs at $z \sim 3$ and found that neutral gas outflow rates exceed the current star formation rate (SFR) of the host galaxy, providing direct observational evidence of ejective AGN feedback. R. L. Davies et al. (2024) have investigated the Na I D absorption from a representative sample at $z \sim 2$ and found that strong neutral gas outflows are widespread in massive galaxies, suggesting that ejective AGN feedback may be the dominant mechanism for quenching at this epoch.

The number density of the massive quiescent population in the early Universe and the emergence of the first quiescent galaxies can place an important constraint on the galaxy formation models (e.g., C. C. Lovell et al. 2023). Recently, JWST has enabled us to explore farther into the early Universe and discover the earliest massive quiescent galaxies. The earliest massive quiescent galaxies currently known and spectroscopically confirmed are at $z \sim 4$–5: $M_\star = 3.8 \times 10^{10} M_\odot$ at $z = 4.657$ (A. C. Carnall et al. 2023a) and $M_\star = 6.3 \times 10^{10} M_\odot$ at $z = 4.53$ (T. Kakimoto et al. 2024), and a low-mass ($M_\star = 4$–$6 \times 10^8 M_\odot$) quiescent galaxy is also found at $z = 7.3$ (T. J. Looser et al. 2024). Recently, K. Glazebrook et al. (2024) have reported one extreme case of a massive quiescent galaxy at $z = 3.2$ that formed half of its mass ($\sim 10^{11} M_\odot$) at $z \gtrsim 11$. As it is still very challenging and expensive to obtain spectra of individual galaxies at such high redshifts, many studies have developed photometric selections to search for quiescent candidates in the early Universe (e.g., S. Alberts et al. 2024; A. S. Long et al. 2024; F. Valentino et al. 2023).

At cosmic noon ($z \sim 2$), when the global SFR and galaxy merger rates peak (e.g., P. Madau & M. Dickinson 2014), galaxy growth and its associated feedback processes are also expected to be the most active. A large fraction of massive galaxies at this epoch appear to be quiescent, based on rest-frame colors (e.g., A. Muzzin et al. 2013). However, obtaining a continuum spectrum for quiescent galaxies at $z \sim 2$ is still very expensive. Another challenge in obtaining IR spectra with ground-based telescopes is heavy atmospheric absorption, making some spectral features undetectable. Only individually targeted samples of quiescent galaxies have been observed for spectroscopy follow-up studies (e.g., S. Belli et al. 2019; M. Kriek et al. 2019; C. D'Eugenio et al. 2020; M. Park et al. 2023), rather than a representative sample, making it difficult to have a complete census of quiescent galaxies at cosmic noon. Many questions about quiescent galaxies at $z \sim 2$ remain unanswered. (1) What are their detailed formation and quenching histories (rapid versus slow quenching and whether they had starbursts before quenching)? (2) What are their physical quenching mechanisms? (3) What are the descendants of the earliest ($z > 3$) quiescent galaxies? The advent of JWST has enabled us to obtain a larger sample of massive quiescent galaxies at $z \sim 2$ (e.g., M. Slob et al. 2024), allowing us to answer these questions.

In this work, we study massive quiescent galaxies selected from a representative sample of 141 galaxies in the Blue Jay survey, a JWST Cycle 1 program. We describe the NIRSpec observations, spectral energy distribution (SED) fitting, and our quiescent sample selection in Section 2. In Section 3, we present the inferred SFHs of our quiescent galaxies and study their quenching histories. In Section 4, we study the neutral and ionized gas properties of our quiescent sample. In Section 5, we predict the number density of massive quiescent galaxies at $z > 3$ based on the SFHs of our entire representative sample. In Section 6, we discuss what the descendants of early massive quiescent galaxies look like at $z \sim 2$. We present a summary and our conclusion in Section 7. Throughout the paper, we adopt the WMAP9 cosmology (G. Hinshaw et al. 2013).

## 2. Data and SED Fitting

### 2.1. The Blue Jay Survey

The Blue Jay survey is a JWST Cycle 1 program (GO 1810; PI: Belli), aiming to study the stellar population and interstellar medium (ISM) of galaxies at cosmic noon. The NIRSpec micro-shutter assembly (MSA) was used to obtain the spectra of 153 galaxies with two pointings in the COSMOS field. The sample was selected from the 3D-HST photometric catalog (R. E. Skelton et al. 2014), considering galaxies in the redshift range $1.7 < z < 3.5$. The goal of the sample selection is to obtain a representative population of galaxies, and the only galaxy properties used for the selection are stellar mass and redshift, to ensure roughly uniform coverage of the $z$, $M$ parameter space. This yields a sample that is representative of the parent population above a stellar mass of $\sim 10^9 M_\odot$ and is not biased in color, SFR, or morphology. For each galaxy, we placed a slitlet with at least two MSA shutters and used a two-point A–B nodding pattern along the slit. To obtain $R \sim 1000$ spectra, we used the medium-resolution gratings (G140M, G235M, and G395M) with exposure times of 13 hr, 3.2 hr, and 1.6 hr, respectively. Through the deep JWST/NIRSpec spectroscopy, we obtained robust continuum detections, thus enabling detailed SED modeling. From the 153 galaxies, we exclude four filler targets at $z \sim 6$ and eight galaxies for which the data reduction failed. Our final Blue Jay sample consists of 141 galaxies. Further details on the Blue Jay observations, data reduction, and sample selection can be found in S. Belli et al. (2024, in preparation).

### 2.2. Prospector SED Fitting

In this section, we describe the SED fitting method used throughout this work. We run Prospector (B. D. Johnson et al. 2021), a fully Bayesian stellar population inference code, to fit both observed photometry and the NIRSpec data. We use broadband photometry from Hubble Space Telescope (HST) and JWST observations only, because of their high-quality calibrations. We measure the photometry for each Blue Jay target in five HST bands—Advanced Camera for Surveys (ACS) and WFC3 instruments—using the mosaics released by the 3D-HST survey (N. A. Grogin et al. 2011; A. M. Koekemoer et al. 2011; R. E. Skelton et al. 2014), and in eight JWST/NIRCam bands, using observations from the PRIMER survey (GO 1837; PI: J. Dunlop). For each target, we define an elliptical aperture in the NIRCam F444W image and then measure the flux and uncertainty within that aperture across all the HST and JWST bands. The measurements account for point-spread function matching and aperture correction and include systematic uncertainty due to contamination from nearby objects or varying local backgrounds. The resulting photometric catalog is publicly available[15] and will be presented in detail in S. Belli et al. (2024, in preparation). Prospector adopts the synthetic stellar population library FSPS (C. Conroy et al. 2009; C. Conroy & J. E. Gunn 2010) to model SEDs. We use the MIST isochrones (J. Choi et al. 2016) and the C3K spectral library (P. A. Cargile et al. 2020) and assume the Chabrier initial mass function (G. Chabrier et al. 2005). The galaxy SED model consists of 29 free parameters describing the contribution of stars, gas, and dust.

The stellar population of galaxies is described as a set of parameters, including redshift, stellar mass, stellar velocity dispersion, metallicity, and a nonparametric SFH (see J. Leja et al. 2019a, 2019b for more details about the setup for nonparametric SFH models). We use 14 SFH bins, starting with the youngest bin of [0, 30 Myr]. The time bins are then equally spaced logarithmically until $0.8 t_{\rm univ}$, where $t_{\rm univ}$ is the age of the Universe, and the two oldest bins are $[0.8 t_{\rm univ}, 0.9 t_{\rm univ}]$ and $[0.9 t_{\rm univ}, 1.0 t_{\rm univ}]$. The purpose of having these two oldest bins is to resolve the star formation at early epochs more accurately, and the overall SFH derived from fitting does not change even if we use the equally spaced bins until the age of the Universe. We fit $\Delta \log \rm SFR$ between two adjacent bins using the Student-T distribution as a prior. We use two different SFH priors in this work. The first is the "continuity" SFH prior, which uses the Student-T distribution with $\sigma = 0.3$ and $\nu = 2.0$ (J. Leja et al. 2019a). This continuity prior is known to disfavor abrupt changes in SFH. We also use the "bursty" prior, introduced in S. Tacchella et al. (2022b), which is a modified version of the continuity prior with $\sigma = 1.0$. The bursty prior does not necessarily favor abrupt changes, but it allows changes more flexibly. The effect of the SFH prior on the inferred SFHs will be discussed in Section 3.2.

The SED model includes dust attenuation, which is modeled assuming two dust components (the birth-cloud component and the diffuse dust component), following S. Charlot & S. M. Fall (2000). Three parameters are used to model dust attenuation: $\tau_{\rm old}$ (optical depth in $V$ band), $n$ (a power-law modifier to the D. Calzetti et al. 2000 law), and extra attenuation toward young stars, $\tau_{\rm young}$ (M. Kriek & C. Conroy 2013). Dust emission is modeled using three parameters ($q_{\rm PAH}$, $\gamma$, and $U_{\rm min}$) describing the IR emission spectrum (B. T. Draine & A. Li 2007). The total amount of energy absorbed by dust is re-emitted in the IR. AGN absorption and emission are not included in our model.

We fit 4000–6700 Å in the rest frame for each galaxy. This wavelength range contains many age-sensitive spectral features, such as Balmer lines, which are important for constraining SFHs. We exclude the region blueward of 4000 Å to avoid interpreting the 4000 Å/Balmer break. While the strength of the break is sensitive to metal abundance, we fit only the total metallicity (leaving the relative abundances fixed to the solar pattern). Also, the polynomial adopted for calibration may affect the shape and strength of the continuum break. We mask out the Na I D line for fitting, as this resonant line can also be absorbed by the intervening neutral gas (see R. L. Davies et al. 2024 for a detailed analysis of the Na I D excess due to neutral gas). For spectrum fitting, we adopt one "jitter" parameter that scales the spectrum uncertainties to obtain a good fit. We add one parameter each for photometry and spectral fitting, which describe the fractions of outliers in photometric points and spectral pixels (see Appendix D in B. D. Johnson et al. 2021 for the outlier model). For the flux calibration and slit-loss corrections, we adopt a polynomial distortion of the spectrum to match the observed (uncalibrated) spectral shape to that of the SED model. We use a polynomial order of 10 for our study and confirm that our main results, including the inferred SFHs, do not change depending on the polynomial order used.

We use three different Prospector setups for our sample. First, we fit only observed photometry for the galaxies where spectral features are not identified (21 galaxies) or for those suffering from undersampling issues (two galaxies). We also do not fit spectra for galaxies whose SEDs cannot be modeled robustly—for example, galaxies undergoing mergers (one galaxy) or containing broad AGN lines (two galaxies). For the remaining 113 galaxies with spectroscopic redshifts, we use different strategies to model emission lines, depending on the possible ionizing sources. For galaxies with emission lines, we fit three Gaussians to [N II] and H$\alpha$ emission lines and measure the line ratio of [N II]/H$\alpha$. Based on the emission-line ratio diagnostics, if the galaxies are thought to be star-forming ([N II]/H$\alpha \ll 1$), we use the CLOUDY prediction (G. J. Ferland et al. 2013; N. Byler et al. 2017) for H$\alpha$ and H$\beta$ lines, as these emission-line luminosities can place constraints on the recent SFR. We marginalize over other emission lines between 4000 and 6700 Å, such as H$\gamma$, H$\delta$, [O III], He I, and [N II]. On the other hand, if the line ratio looks more consistent with AGN diagnostics ([N II]/H$\alpha \gtrsim 1$), we marginalize over all emission lines, including H$\alpha$ and H$\beta$, to avoid misinterpreting the ionizing sources. In the presence of broad AGN lines, we mask out all the broad emission lines.

---
[15] On Zenodo doi:10.5281/zenodo.13292819.

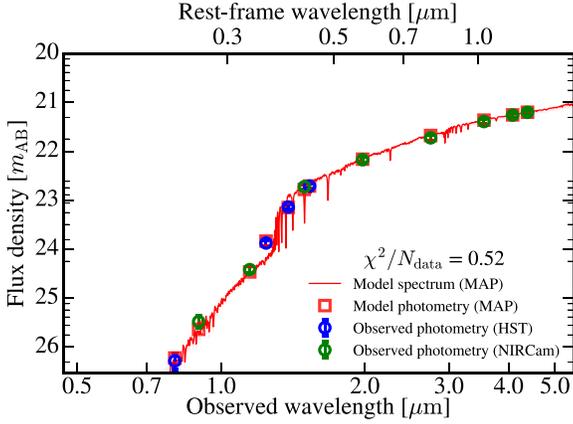
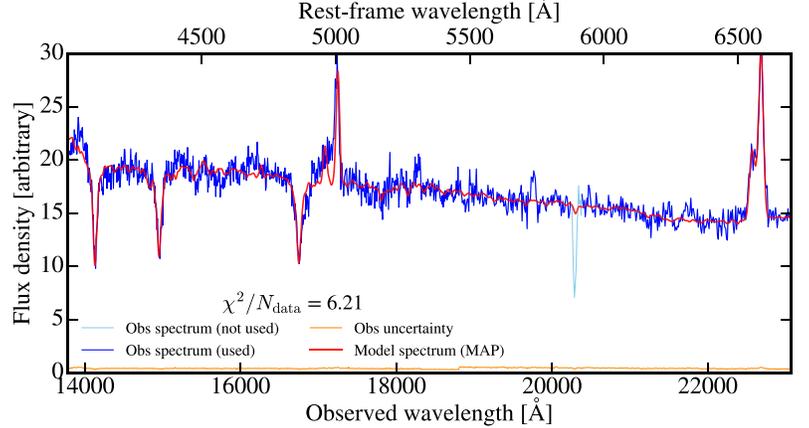
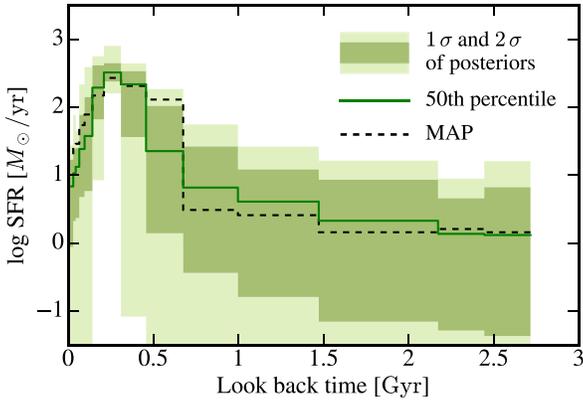
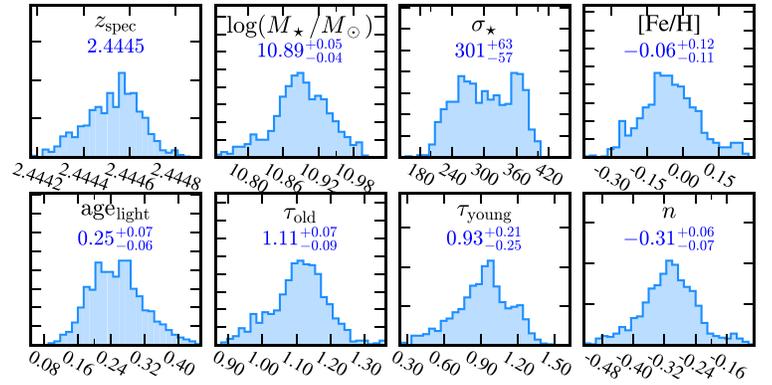

**Figure 1.** Prospector fitting results of an example Blue Jay galaxy, COSMOS-11142. (a) Observed photometry is shown with circles (blue: HST/ACS+WFC3; green: JWST/NIRCam). The best-fit SED model is shown in red. (b) NIRSpec data (in arbitrary units) is shown as a blue line, and the best-fit spectrum is again shown as a red line. (c) Nonparametric SFH derived from fitting using the continuity SFH prior. The green and yellow–green shades include $1\sigma$ and $2\sigma$ of the posteriors at each SFH time bin. The green line shows the 50th percentile of the posteriors at each time bin, and the dashed black line shows the SFH from MAP. (d) Posterior distribution of the stellar and dust parameters: (1) redshift ($z_{\rm spec}$); (2) stellar mass ($\log(M_\star/M_\odot)$); (3) stellar velocity dispersion ($\sigma_\star$); (4) metallicity ([Fe/H]); (5) light-weighted stellar ages (age$_{\rm light}$); (6) optical depth for the diffuse dust component ($\tau_{\rm old}$); (7) optical depth for the birth-cloud component ($\tau_{\rm young}$); and (8) the power-law modifier to the shape of the D. Calzetti et al. (2000) dust attenuation curve ($n$).

Figure 1 shows the Prospector fitting results of an example Blue Jay galaxy, COSMOS-11142 (the same galaxy studied in S. Belli et al. 2024), fitting both observed photometry and the NIRspec spectrum. Panel (a) shows the photometric data we used with circles (blue: HST/ACS+WFC3; green: JWST/NIRCam). The red line represents the best-fit SED model or maximum a posteriori (MAP). The NIRspec $R \sim 1000$ spectrum (in arbitrary units) is shown as the blue line in panel (b). Note that the Na I D line is masked out (shown as a sky-blue line), as this line can also be absorbed by intervening gas.

The red line in panel (b) also shows the (uncalibrated) best-fit model spectrum, and the green line is the polynomial that must be multiplied by the uncalibrated spectrum to obtain the calibrated spectrum in panel (a). Panel (c) shows the derived nonparametric SFH using the continuity SFH prior. Panel (d) shows the posterior distribution of fitted stellar and dust parameters. We summarize all the derived stellar and dust parameters for our 14 quiescent galaxies in Table 1.

### 2.3. Quiescent Sample

In this section, we describe how we select our quiescent sample. Figure 2(a) shows the SFR averaged over the most recent 30 Myr (the youngest SFH bin used) as a function of stellar mass for all our 141 Blue Jay galaxies. The error bar shows the 16th and 84th percentiles of the SFR and stellar mass derived from the Prospector fitting. Galaxies for which only observed photometry is fitted are shown as crosses, while the circles show the galaxies where we fit both observed photometry and the NIRSpec data. The color code shows the light-weighted (using the bolometric luminosity as weight)

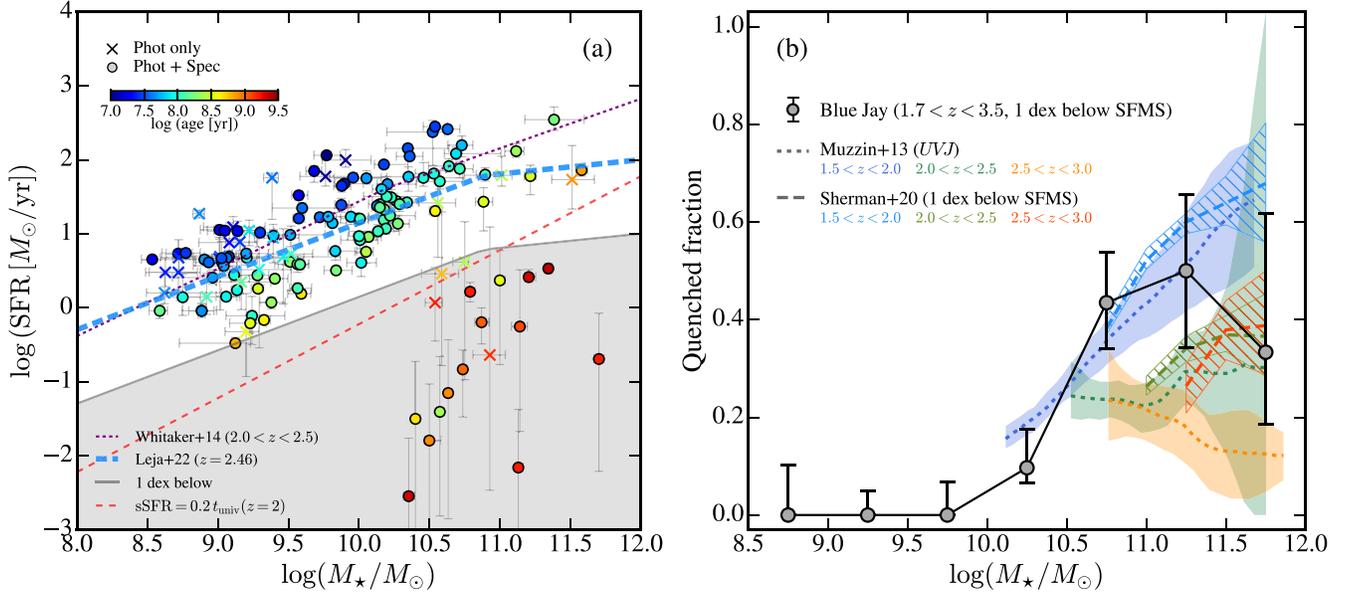

**Figure 2.** (a) SFR averaged over the most recent 30 Myr shown as a function of galaxy stellar mass. Galaxies are color-coded by light-weighted ages. The error bar indicates the 16th and 84th percentiles of the SFR. The different symbols indicate the different Prospector setups used: fitting only photometry (crosses) and fitting both photometry and NIRSpec data (circles). The purple dotted line shows the SFMS at $z \sim 2$ from K. E. Whitaker et al. (2014), and the blue dashed line shows the SFMS at $z \sim 2.46$ (median redshift of the Blue Jay sample) from J. Leja et al. (2022). We select 14 quiescent galaxies located 1 dex below the SFMS from J. Leja et al. (2022) at their respective redshifts (the gray shaded region shows 1 dex below the SFMS at $z = 2.46$ for a visual guide). Most of our quiescent galaxies also satisfy sSFR $< 0.2/t_{univ}(z)$, which is shown as the red dashed line. (b) Fraction of quiescent galaxies as a function of galaxy stellar mass. The error bar includes the $1\sigma$ confidence interval from the binomial distribution (e.g., E. Cameron 2011). We find that the mass trend of the quenched fractions in the Blue Jay sample is consistent with other observations of large photometric samples at similar redshifts.

**Table 1**  
Prospector Fitting Results of 14 Massive Quiescent Galaxies in Our Sample

| (1) 3D-HSTCOSMOS ID | (2) $z_{spec}$ | (3) $\log(M_\star/M_\odot)$ | (4) [Fe/H] | (5) $\hat{\tau}_{dust,old}$ | (6) $\hat{\tau}_{dust,young}$ | (7) $n$ |
|---|---|---|---|---|---|---|
| 7549  | 2.626 | $10.78^{+0.07}_{-0.06}$ | $-0.62^{+0.20}_{-0.16}$ | $0.85^{+0.17}_{-0.10}$ | $0.93^{+0.30}_{-0.23}$ | $0.03^{+0.15}_{-0.12}$ |
| 21477 | 2.473 | $10.63^{+0.03}_{-0.03}$ | $-0.10^{+0.15}_{-0.18}$ | $0.62^{+0.12}_{-0.16}$ | $0.78^{+0.21}_{-0.31}$ | $0.14^{+0.10}_{-0.13}$ |
| 11142 | 2.444 | $10.89^{+0.05}_{-0.04}$ | $-0.06^{+0.12}_{-0.11}$ | $1.11^{+0.07}_{-0.09}$ | $0.93^{+0.21}_{-0.25}$ | $-0.31^{+0.06}_{-0.07}$ |
| 10565 | 2.442 | $10.88^{+0.06}_{-0.05}$ | $-0.23^{+0.09}_{-0.09}$ | $0.76^{+0.10}_{-0.09}$ | $0.69^{+0.20}_{-0.14}$ | $0.10^{+0.16}_{-0.15}$ |
| 10339 | 2.363 | $10.42^{+0.03}_{-0.04}$ | $-0.31^{+0.17}_{-0.31}$ | $0.63^{+0.11}_{-0.10}$ | $0.56^{+0.11}_{-0.15}$ | $-0.54^{+0.23}_{-0.16}$ |
| 9395  | 2.127 | $10.74^{+0.03}_{-0.02}$ | $-0.10^{+0.06}_{-0.08}$ | $0.37^{+0.06}_{-0.05}$ | $0.33^{+0.08}_{-0.06}$ | $-0.52^{+0.20}_{-0.14}$ |
| 10400 | 2.099 | $10.33^{+0.04}_{-0.03}$ | $0.03^{+0.16}_{-0.17}$ | $0.07^{+0.09}_{-0.05}$ | $0.06^{+0.09}_{-0.05}$ | $-0.11^{+0.16}_{-0.19}$ |
| 11494 | 2.091 | $11.70^{+0.03}_{-0.03}$ | $0.09^{+0.06}_{-0.05}$ | $0.44^{+0.06}_{-0.06}$ | $0.51^{+0.13}_{-0.11}$ | $0.07^{+0.12}_{-0.18}$ |
| 18668 | 2.086 | $11.13^{+0.06}_{-0.21}$ | $-0.07^{+0.16}_{-0.14}$ | $1.34^{+0.11}_{-0.11}$ | $1.39^{+0.31}_{-0.26}$ | $0.12^{+0.16}_{-0.35}$ |
| 16419 | 1.925 | $11.32^{+0.03}_{-0.02}$ | $-0.03^{+0.03}_{-0.02}$ | $0.44^{+0.05}_{-0.04}$ | $0.45^{+0.06}_{-0.08}$ | $0.04^{+0.15}_{-0.13}$ |
| 8469  | 1.868 | $10.58^{+0.07}_{-0.04}$ | $-0.21^{+0.13}_{-0.12}$ | $0.50^{+0.08}_{-0.09}$ | $0.52^{+0.14}_{-0.12}$ | $0.17^{+0.14}_{-0.15}$ |
| 10128 | 1.852 | $11.25^{+0.05}_{-0.05}$ | $-0.48^{+0.10}_{-0.08}$ | $0.83^{+0.07}_{-0.08}$ | $0.92^{+0.22}_{-0.21}$ | $0.24^{+0.10}_{-0.09}$ |
| 10592 | 1.801 | $11.17^{+0.02}_{-0.02}$ | $0.19^{+0.08}_{-0.10}$ | $0.26^{+0.08}_{-0.08}$ | $0.24^{+0.11}_{-0.08}$ | $0.13^{+0.16}_{-0.19}$ |
| 8013  | 1.690 | $10.56^{+0.04}_{-0.04}$ | $-0.07^{+0.07}_{-0.16}$ | $0.69^{+0.08}_{-0.07}$ | $0.70^{+0.11}_{-0.11}$ | $-0.32^{+0.09}_{-0.09}s$ |

**Note.** Column (1): 3D-HST COSMOS ID. Column (2): fitted spectroscopic redshift. Column (3): stellar mass of galaxy. Column (4): metallicity. Column (5): optical depth (in V band) for the diffuse dust component (see the details in C. Conroy et al. 2009). Column (6): optical depth for the birth-cloud component. Column (7): the power-law modifier to the shape of the D. Calzetti et al. (2000) dust attenuation curve (see the details in M. Kriek & C. Conroy 2013). The fitting results here are based on using the continuity SFH prior. The results using bursty prior can be found in Table A1.

ages of galaxies. The purple dotted line shows the star-forming main sequence (SFMS) at $2.0 < z < 2.5$ from K. E. Whitaker et al. (2014), and the blue dashed line shows the SFMS at $z = 2.46$ (median redshift of the Blue Jay sample) from J. Leja et al. (2022).

In this study, we select 14 quiescent galaxies as the galaxies with specific SFR (sSFR) 1 dex below the SFMS from J. Leja et al. (2022) at their respective redshifts (shown as the gray shaded region). We remove galaxies that satisfy this sSFR criterion but have no spectroscopic redshift (COSMOS-12332, 17713, 20195, and 21541). The majority of our quiescent sample would also be identified as quiescent by the rest-frame UVJ colors (see Appendix A and Figure A1); only COSMOS-8469 and 18668 are outside the quiescent region but near its border. The red dashed line in Figure 2(a) indicates sSFR = $0.2/t_{univ}(z)$ at $z = 2$. This sSFR cut has been widely used to select quiescent galaxies in many other studies (e.g., A. Gallazzi et al. 2014; C. Pacifici et al. 2016; S. Tacchella et al. 2022a; A. C. Carnall et al. 2023b). The figure shows that the quiescent sample also satisfies this sSFR cut (sSFR $< 0.2/t_{univ}(z)$). In Appendix B, we present the NIRSpec data and the best-fit SED models from the Prospector fitting for all our 14 quiescent galaxies (see Figures A2, A3, and A4).

Figure 2(b) shows the fraction of quiescent galaxies as a function of their stellar mass. All our quiescent galaxies are massive ($\log(M_\star/M_\odot) > 10.0$), and the fraction increases steeply with stellar mass. The dotted lines (and shades) are the observed quenched fractions (and $1\sigma$ uncertainties) at $1.5 < z < 2.0$ (blue), $2.0 < z < 2.5$ (green), and $2.5 < z < 3.0$ (orange) from A. Muzzin et al. (2013). In their study, quiescent galaxies are identified by their rest-frame $UVJ$ colors. The dashed lines (and hatched regions) show the observed quenched fractions (and $1\sigma$ uncertainties) from S. Sherman et al. (2020; the same color code used for different redshift bins). They defined quiescent galaxies as galaxies with sSFR more than 1 dex below the SFMS at the respective redshift. In the Blue Jay sample, 23% of the massive galaxies ($\log(M_\star/M_\odot) > 10$) are quiescent. The median redshift of the entire Blue Jay sample is $z = 2.46$, and for the massive quiescent galaxies, the median redshift is $z = 2.09$. The mass trend of the quenched fraction in the Blue Jay sample is consistent with observations of large photometric samples at similar redshifts.

## 3. SFH

### 3.1. Formation and Quenching Histories at $z \sim 2$

We fit both observed photometry and the NIRSpec data using Prospector to infer the SFHs of the 14 massive quiescent galaxies. For consistency, we use the emission-line marginalization model for all emission lines between 4000 and 6700 Å in the rest frame. To study the formation and quenching history in more detail, we define the formation and quenching epochs of a galaxy. The formation epoch ($z_{form}$ in redshift and $t_{form}^{lbt}$ in lookback time in Gyr) is defined as the epoch when 50% of the final (surviving) stellar mass was formed. The quenching epoch is defined as the epoch when sSFR drops below sSFR $< 0.2\, t_{univ}(z)$, where $t_{univ}(z)$ is the age of the Universe (used in many other studies; e.g., A. Gallazzi et al. 2014; C. Pacifici et al. 2016; A. C. Carnall et al. 2023b). In the case of multiple quenching events (which is mostly relevant when using the bursty prior; see Section 3.2), we first identify two quenching epochs: the first quenching (the earliest quenching event since formation) and the second quenching (the most recent quenching event). We then define the "main" quenching event by comparing the star formation peaks associated with these two quenching events. If, after the first quenching, the SFR rejuvenates to >10% of the star formation peak before the first quenching, the second quenching event is considered the main quenching. Otherwise, we consider the first quenching the main quenching.

Based on the formation and quenching histories, we classify our 14 massive quiescent galaxies into three groups:

1. Group-1: early formation ($t_{form}^{lbt} > 1.0$ Gyr) and early quenching ($t_{quench}^{lbt} > 0.5$ Gyr);

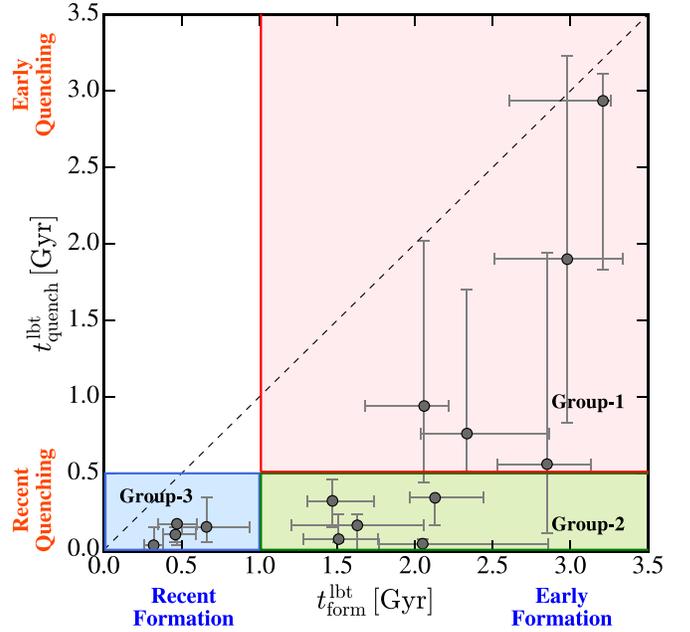

**Figure 3.** Fourteen Blue Jay massive quiescent galaxies in the plane of formation ($t_{form}^{lbt}$) and quenching epochs ($t_{quench}^{lbt}$), both in lookback time in Gyr. We classify them into three groups based on these two epochs. Group-1 (red region): galaxies that are early formed and early quenched. Group-2 (green region): galaxies that are early formed and recently quenched. Group-3 (blue region): galaxies that are recently formed and recently quenched. We identify five galaxies in Group-1, five in Group-2, and four in Group-3.

2. Group-2: early formation ($t_{form}^{lbt} > 1.0$ Gyr) and recent quenching ($t_{quench}^{lbt} < 0.5$ Gyr); and
3. Group-3: recent formation ($t_{form}^{lbt} < 1.0$ Gyr) and recent quenching ($t_{quench}^{lbt} < 0.5$ Gyr).

Figure 3 shows 14 Blue Jay massive quiescent galaxies in the plane of the formation ($t_{form}^{lbt}$) and quenching epochs ($t_{form}^{lbt}$) and how they are divided into the three groups. Note that the exact thresholds of $t_{form} = 1.0$ Gyr and $t_{quench} = 0.5$ Gyr are arbitrary and depend on the exact definition of the "quenching epoch." However, these cuts are useful to identify three (qualitatively) distinct types of SFHs: (1) old galaxies quenched at early epochs; (2) recently quenched galaxies after steady formation; and (3) recently quenched galaxies after a major starburst.

Figure 4 shows the SFHs of the 14 massive quiescent galaxies observed in the Blue Jay sample, inferred from the Prospector fitting using the continuity SFH prior. In each group (in each column in Figure 4), galaxies are in descending order of observed redshift, which is shown as the black vertical line (and the region is shaded gray beyond the observed epoch). The 3D-HST COSMOS IDs and the light-weighted age of each galaxy are shown at the right corner of each plot. In each SFH panel, the dashed line is the SFH from the MAP, and the solid line connects the median SFR of the posteriors in each time bin. The green and yellow–green shades include $1\sigma$ and $2\sigma$ posterior distributions, respectively. The blue arrow in each panel shows the formation epoch, with the sky-blue bar indicating the 16th to the 84th percentiles of the formation redshift. The orange arrow shows the quenching epoch.

We identify five galaxies as Group-1 galaxies. They are the oldest quiescent galaxies in our sample, with light-weighted ages of age$_{light} \approx 2.0$ Gyr. Five galaxies are classified as Group-2. These galaxies appear to have steady formation, with nearly

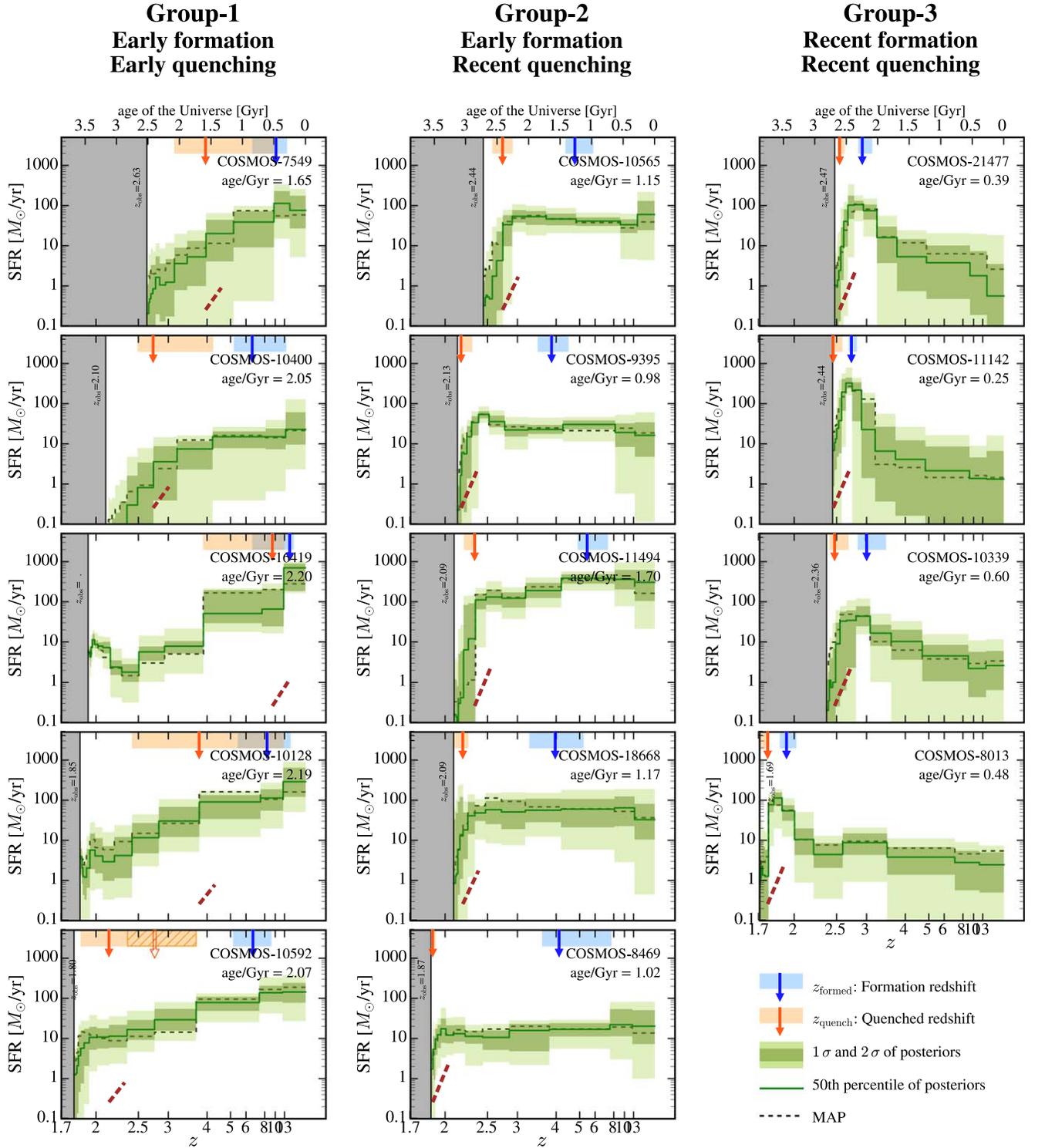

**Figure 4.** SFHs of 14 massive quiescent galaxies grouped into Group-1: early formation and early quenching (left column); Group-2: early formation and recent quenching (middle column); and Group-3: recent formation and recent quenching (right column). The green and yellow–green shades include $1\sigma$ and $2\sigma$ of the posteriors at each SFH time bin, and the green line shows the 50th percentile of the posteriors at each time bin. The dashed black line shows the SFH from MAP. The blue and orange arrows indicate the formation ($z_{form}$) and quenching epoch ($z_{quench}$), and the sky-blue and orange bars show the 16th to 84th percentiles of $z_{form}$ and $z_{quench}$, respectively. The red dashed slope shows the quenching speed, $v_{quench}$ (see the text for the definition). The 3D-HST IDs and light-weighted stellar age are shown at the upper right corner of each panel. We classify 14 massive quiescent galaxies into three groups based on their formation and quenching histories (early/recent formation and early/recent quenching) and identify five galaxies in Group-1, five galaxies in Group-2, and four galaxies in Group-3.

constant SFR over their lifetime until recently. They formed half of their stellar mass at early epochs and have recently experienced abrupt quenching. They have intermediate ages in our sample ($age_{light} \approx 1.0$ Gyr). Last, four galaxies are classified as Group-3. These galaxies formed very recently, with intense starbursts followed by rapid quenching. These

Table 2
Properties of 14 Blue Jay Massive Quiescent Galaxies Grouped by Their Star Formation and Quenching Histories

| (1) ID | (2) 3D-HST COSMOS ID | (3) $z_{\rm obs}$ | (4) $\log(M_\star/M_\odot)$ | (5) $\mathrm{age_{light}}$ (Gyr) | (6) $z_{\rm form}$ | (7) $t_{\rm form}^{\rm lbt}$ (Gyr) | (8) $z_{\rm quench}$ | (9) $t_{\rm quench}^{\rm lbt}$ (Gyr) | (10) $v_{\rm quench}$ (dex Gyr$^{-1}$) |
|---|---|---|---|---|---|---|---|---|---|
| Group-1: Early Formation and Early Quenching | | | | | | | | | |
| 1 | 7549 | 2.627 | 10.78 | $1.65^{+0.37}_{-0.31}$ | $10.32^{+3.73}_{-3.75}$ | $2.06^{+0.16}_{-0.38}$ | $3.97^{+5.74}_{-0.84}$ | $0.94^{+1.08}_{-0.50}$ | $2.54^{+2.38}_{-1.25}$ |
| 2 | 10400 | 2.099 | 10.33 | $2.05^{+0.39}_{-0.31}$ | $6.58^{+7.08}_{-1.37}$ | $2.33^{+0.53}_{-0.29}$ | $2.73^{+1.47}_{-0.24}$ | $0.76^{+0.94}_{-0.25}$ | $2.36^{+2.80}_{-1.21}$ |
| 3 | 16419 | 1.925 | 11.32 | $2.20^{+0.15}_{-0.24}$ | $16.36^{+2.89}_{-9.81}$ | $3.21^{+0.05}_{-0.60}$ | $9.46^{+3.33}_{-5.58}$ | $2.93^{+0.18}_{-1.10}$ | $3.00^{+0.37}_{-0.40}$ |
| 4 | 10128 | 1.852 | 11.25 | $2.19^{+0.27}_{-0.26}$ | $8.46^{+7.94}_{-3.02}$ | $2.98^{+0.36}_{-0.47}$ | $3.76^{+8.78}_{-1.35}$ | $1.90^{+1.33}_{-1.07}$ | $2.03^{+1.26}_{-0.87}$ |
| 5 | 10592 | 1.801 | 11.17 | $2.07^{+0.26}_{-0.26}$ | $6.65^{+2.44}_{-1.49}$ | $2.85^{+0.28}_{-0.32}$ | $2.14^{+1.52}_{-0.28}$ | $0.56^{+1.38}_{-0.45}$ | $2.07^{+5.13}_{-1.11}$ |
| Group-2: Early Formation and Recent Quenching | | | | | | | | | |
| 6 | 10565 | 2.442 | 10.88 | $1.15^{+0.13}_{-0.13}$ | $4.81^{+1.02}_{-0.45}$ | $1.47^{+0.27}_{-0.16}$ | $2.75^{+0.16}_{-0.17}$ | $0.32^{+0.14}_{-0.17}$ | $5.67^{+3.88}_{-2.12}$ |
| 7 | 9395 | 2.127 | 10.74 | $0.98^{+0.15}_{-0.15}$ | $3.88^{+0.58}_{-0.42}$ | $1.51^{+0.25}_{-0.23}$ | $2.18^{+0.16}_{-0.01}$ | $0.07^{+0.16}_{-0.02}$ | $7.13^{+3.79}_{-1.92}$ |
| 8 | 11494 | 2.091 | 11.70 | $1.70^{+0.14}_{-0.12}$ | $5.51^{+1.70}_{-0.59}$ | $2.13^{+0.31}_{-0.16}$ | $2.34^{+0.00}_{-0.14}$ | $0.34^{+0.00}_{-0.18}$ | $7.60^{+6.53}_{-3.44}$ |
| 9 | 18668 | 2.086 | 11.13 | $1.17^{+0.26}_{-0.88}$ | $4.00^{+1.20}_{-0.74}$ | $1.63^{+0.43}_{-0.42}$ | $2.20^{+0.05}_{-0.11}$ | $0.16^{+0.07}_{-0.16}$ | $5.82^{+4.30}_{-2.18}$ |
| 10 | 8469 | 1.868 | 10.58 | $1.02^{+0.21}_{-0.25}$ | $4.14^{+3.47}_{-0.56}$ | $2.05^{+0.81}_{-0.28}$ | $1.89^{+0.02}_{-0.01}$ | $0.04^{+0.03}_{-0.01}$ | $7.85^{+12.52}_{-4.45}$ |
| Group-3: Recent Formation and Recent Quenching | | | | | | | | | |
| 11 | 21477 | 2.474 | 10.63 | $0.39^{+0.13}_{-0.08}$ | $2.94^{+0.16}_{-0.09}$ | $0.46^{+0.13}_{-0.08}$ | $2.56^{+0.05}_{-0.05}$ | $0.10^{+0.05}_{-0.05}$ | $8.22^{+4.76}_{-2.26}$ |
| 12 | 11142 | 2.444 | 10.89 | $0.25^{+0.07}_{-0.06}$ | $2.75^{+0.06}_{-0.06}$ | $0.32^{+0.06}_{-0.06}$ | $2.47^{+0.11}_{-0.03}$ | $0.03^{+0.12}_{-0.03}$ | $7.22^{+5.58}_{-1.95}$ |
| 13 | 10339 | 2.363 | 10.42 | $0.60^{+0.22}_{-0.18}$ | $3.03^{+0.40}_{-0.19}$ | $0.66^{+0.28}_{-0.16}$ | $2.49^{+0.18}_{-0.09}$ | $0.15^{+0.19}_{-0.10}$ | $7.32^{+5.44}_{-2.25}$ |
| 14 | 8013 | 1.690 | 10.56 | $0.48^{+0.13}_{-0.14}$ | $1.94^{+0.08}_{-0.07}$ | $0.47^{+0.13}_{-0.12}$ | $1.77^{+0.00}_{-0.07}$ | $0.17^{+0.00}_{-0.14}$ | $7.58^{+5.99}_{-1.46}$ |

**Note.** Column (1): ID. Column (2): 3D-HST COSMOS ID. Column (3): observed redshift. Column (4): stellar mass. Column (5): light-weighted stellar age (in Gyr). Column (6): formation redshift. Column (7): formation epoch in lookback time (in Gyr). Column (8): quenching redshift. Column (9): quenching epoch in lookback time (in Gyr). Column (10): quenching speed (dex Gyr$^{-1}$). All the results are based on the Prospector fitting using the continuity prior. The results using the bursty prior can be found in Table A2.

PSBs are therefore dominated by young stellar populations, so they have the youngest ages in our quiescent sample ($\mathrm{age_{light}} \approx 0.5$ Gyr).

To measure how rapidly/slowly galaxies are quenched, we define the quenching speed ($v_{\rm quench}$) as follows:

$$v_{\rm quench}\ [{\rm dex/Gyr}] = \frac{\log{\rm SFR}(t_{\rm peak}) - \log{\rm SFR}(t_{\rm lowest})}{t_{\rm peak} - t_{\rm lowest}}. \quad (1)$$

The $t_{\rm peak}$ and $t_{\rm lowest}$ are the epochs within the time window of [$t_{\rm quench} - 0.25$ Gyr, $t_{\rm quench} + 0.25$ Gyr], where the SFRs are at their peak and their lowest point, respectively. We measure how quickly logSFR drops between $t_{\rm peak}$ and $t_{\rm lowest}$. In Figure 4, we present this quenching speed (slope) as a red dashed slope in each SFH panel. The quenching speed for each galaxy is also given in Table 2.

All the recently quenched galaxies (in Group-2 and Group-3) are rapidly quenched with $v_{\rm quench} \gtrsim 6$ dex Gyr$^{-1}$. Note that the SFHs presented in Figure 4 are derived using the continuity SFH prior. The continuity prior is designed to disfavor abrupt changes in SFH, such as rapid quenching, and yet most of our quiescent galaxies appear to be quenched very rapidly at recent epochs. The inferred quenching speeds for galaxies quenched at early epochs (Group-1) are lower than those of recently quenched galaxies. However, this might be a biased result for the following reasons: (1) the SFH time bins are coarser at older ages; (2) any sharp transition in SFH at old ages is hard to recover, as inferring the exact ages of old stellar populations is very challenging; and (3) the accretion of smaller galaxies after quenching can smooth out the SFH, diluting the rapid quenching signal in the past. Therefore, the quenching speeds inferred for early-quenched galaxies should be considered lower limits. Galaxies quenched at early epochs are likely quenched rapidly as well, but we cannot estimate precisely how rapidly they were quenched in the past.

### 3.2. Dependence on SFH Prior

Several studies have shown that high-$z$ galaxies tend to have bursty star formation. Especially, lower-mass galaxies seem to go through numerous mini-bursts and mini-quenching (e.g., T. J. Looser et al. 2023; T. Dome et al. 2024). However, the continuity SFH prior we used in Figure 4 disfavors abrupt changes in SFH, including starbursts, rapid quenching, and rejuvenation, which all seem to be more frequent at higher redshifts. Indeed, the continuity prior does not seem to be able to reproduce the past burst events, as shown in the test we did with mock data (see Appendix D for more details). Thus, many works studying high-$z$ galaxies have instead used other SFH priors that can allow more flexible changes in SFH (e.g., K. A. Suess et al. 2022; S. Tacchella et al. 2022b, 2023; B. Wang et al. 2023; T. Harvey et al. 2024; D. Narayanan et al. 2024). To see how the derived SFH changes with different priors, we refit all our 14 quiescent galaxies with the bursty SFH prior (see Section 2.2 for the details of the bursty prior).

Figure 5 shows the SFHs of the same 14 quiescent galaxies (in the same format), but derived from the Prospector fitting using the bursty SFH prior. The Prospector fitting results using the bursty prior are given in Table A1. We also measure the formation ($z_{\rm form}$) and quenching epochs ($z_{\rm quench}$) and the quenching speed ($v_{\rm quench}$) in the same way that we did with the SFHs derived using the continuity prior. We summarize all our measurements for these epochs in Table A2. In Figure A5, we compare $z_{\rm form}$, $z_{\rm quench}$, and light-weighted ages derived using the continuity and bursty SFH priors. We find that, broadly, all these parameters derived from the two priors are consistent.

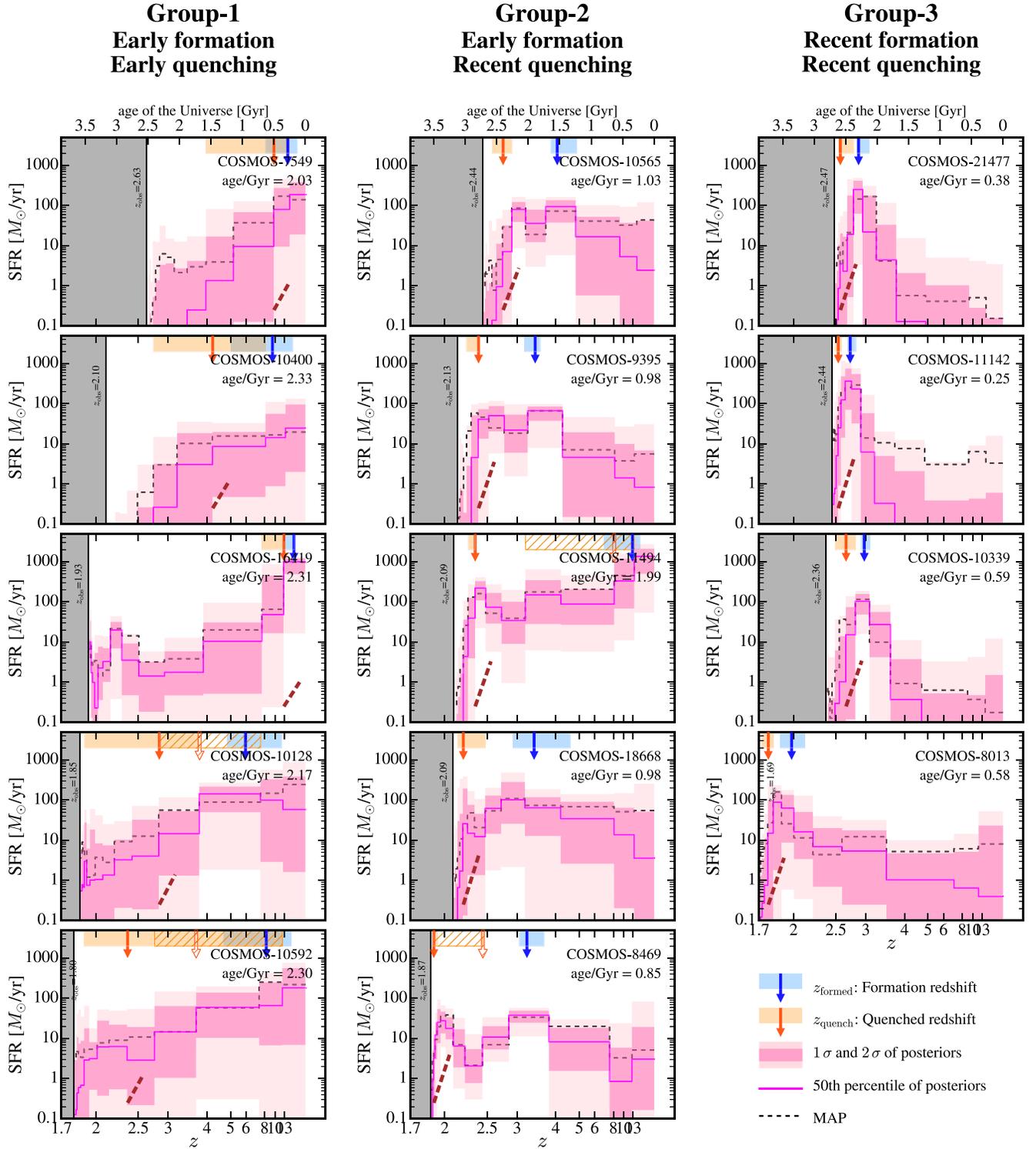

**Figure 5.** SFHs of the same 14 massive quiescent galaxies derived using the bursty SFH prior, in the same format as Figure 4. The empty orange arrow and the hatched bar indicate the first (earliest) quenching epoch (the median and 16th to 84th percentiles) that happened more than 1 Gyr before the "main" quenching event (see the text in Section 3 for how the first/second and main quenching events are defined). We find that the SFHs overall look consistent for galaxies in Group-1 and Group-3. However, the galaxies in Group-2, with stellar ages of ~1 Gyr, show different SFHs when different SFH priors are used: steady formation with nearly constant SFRs until recent quenching (continuity prior) vs. repeated bursts and rapid quenching (bursty prior). Broadly, the formation and quenching epochs and the stellar ages derived from continuity and bursty priors are consistent (see also Figure A5).

Therefore, the group classification (early versus recent formation and early versus recent quenching) remains valid even when the bursty prior is used.

Overall, the inferred SFHs look consistent for galaxies in Group-1 and Group-3. For early quenched galaxies in Group-1, some of them (e.g., COSMOS-7549, 10400, and 16419) are quenched more abruptly at earlier epochs and show very low SFRs since quenching. COSMOS-16419 appears to have experienced several rejuvenation events since its first quenching. The SFHs inferred for the galaxies in Group-3 also look consistent; they all experience recent intense starbursts followed by rapid quenching. When the bursty prior is used, they tend to be quenched with even higher quenching speeds ($v_{\rm quench} > 12 \text{ dex Gyr}^{-1}$). This is because the quenching speed measurement is closely related to the parameter in the model (which is the ratio between log SFR between the two adjacent time bins) and is therefore sensitive to how this parameter is sampled with the assumed prior. Thus, the quenching speed inferred using the continuity prior is likely a lower limit.

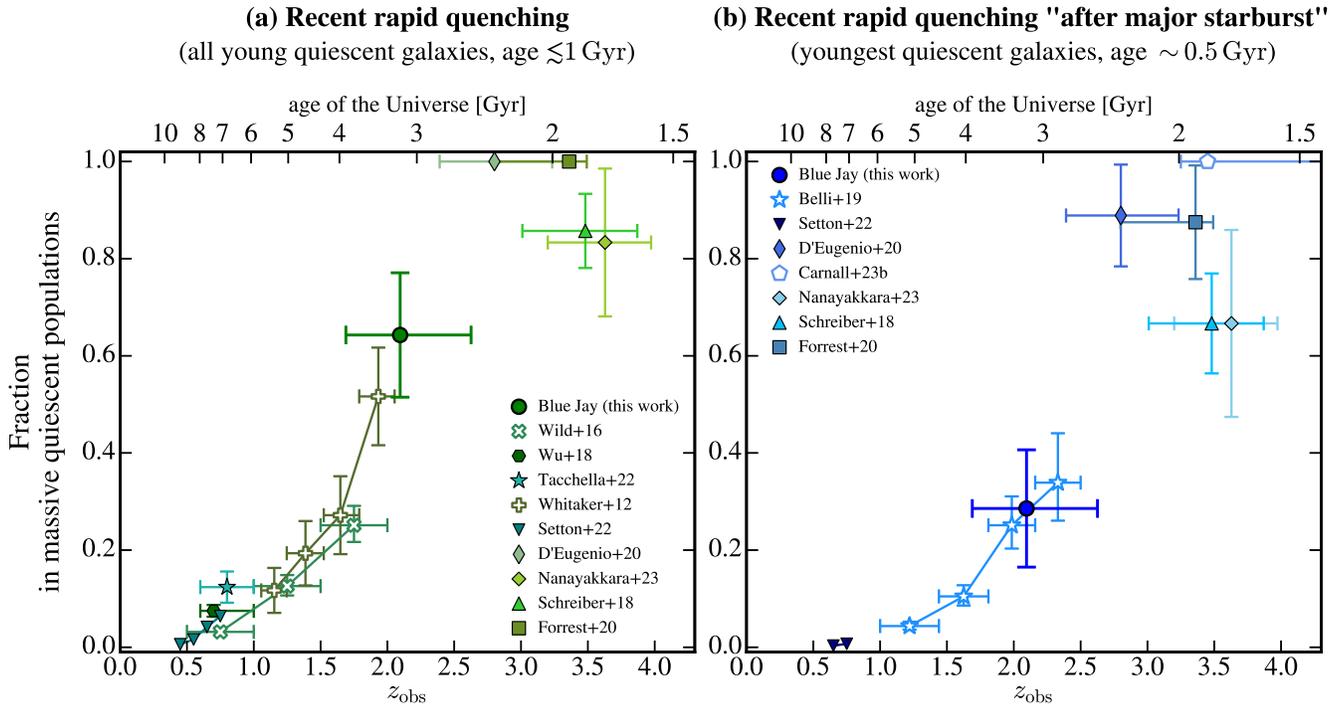

**Figure 6.** Fraction of quiescent galaxies that are rapidly quenched (left) and PSBs (right) among massive ($\log(M_\star/M_\odot) > 10$) quiescent populations as a function of redshift. The RQs include PSBs. The green and blue circles show the fractions from the Blue Jay quiescent sample. The fractions at different redshifts from the literature are also shown with different symbols (summarized in Table 3). We find that approximately 60% (9/14) of our Blue Jay quiescent galaxies are rapidly quenched, and only 30% (4/14) are rapidly quenched after "major" starbursts (PSBs). These fractions at $z \sim 2$ agree well with the trend where the fraction of young quiescent galaxies increases with redshift.

Galaxies in Group-2 (early formation and recent quenching), with intermediate stellar ages of $\sim 1$ Gyr, show different SFHs when different priors are used. When the continuity prior ($\sigma = 0.3$, where $\sigma$ is the width in the Student-T distribution) is used, all of them appear to have very steady formation, with nearly constant SFRs throughout cosmic time, and experience recent abrupt quenching. However, when the bursty prior ($\sigma = 1.0$) is used, they are fitted with the SFH having multiple bursts and quenching in the past. We also test by trying different $\sigma$ values and find a trend; galaxies with intermediate ages are fitted with SFH with more bursts and quenching as we use an SFH prior with higher $\sigma$.

It seems that both scenarios (steady formation and recent quenching versus repeated bursts and quenching) fit our data equally well (best-fit SEDs having comparable $\chi^2$ values), so we cannot distinguish between these two SFHs for galaxies with intermediate ages. It is tempting to trust the bursty prior over the continuity prior, because the bursty prior is more flexible by design and is even able to recover past starburst events (as tested with mock data). However, we should not overinterpret the results when the bursty prior is used. Also, the bursty prior will more likely lean into the outshining effects, which can cause a systematic decrease in the stellar ages (e.g., B. Wang et al. 2024). However, for our quiescent sample, we did not see any systematic differences in the light-weighted stellar ages or any other derived parameters and measured epochs (Figure A5). Throughout the paper, we use SFHs derived using both continuity and bursty priors and give the predicted number density at high redshifts based on these two priors (see Section 5).

### 3.3. Fraction of RQ versus PSB Galaxies

In this section, we measure the fraction of RQs and PSBs from our Blue Jay quiescent sample to answer: (1) how many galaxies are rapidly/slowly quenched at $z \sim 2$; and (2) how many of the recently quenched galaxies are PSBs. We define PSBs as galaxies that experienced "major" starbursts before rapid quenching, such that galaxies are basically "formed" from these starburst events. In our Blue Jay quiescent sample, we call galaxies in Group-3—recent formation and recent quenching—PSBs. The RQs are more inclusive populations. In Section 3.1, we show that all the recently quenched galaxies (in Group-2 and Group-3) are rapidly quenched with $v_{\rm quench} \gtrsim 6 \text{ dex Gyr}^{-1}$. Thus, the RQs include PSBs (Group-3) and the galaxies in Group-2 that are rapidly quenched after a steady (as the continuity prior predicts) or bursty (as revealed by using the bursty prior) formation.

Figure 6 shows the fraction of (a) all RQs and (b) galaxies rapidly quenched after a major starburst (PSBs) as a function of redshift. The fractions from the Blue Jay quiescent galaxies are shown as the green (left) and blue (right) circles.

To compare the fractions at different redshifts, we reclassify young quiescent galaxies from the literature into RQs (inclusive) or PSBs. Young quiescent galaxies are often selected using the strength of Balmer absorption (e.g., $H\delta > 5$ Å used in P.-F. Wu et al. 2018) or colors (e.g., K. E. Whitaker et al. 2012b; V. Wild et al. 2016). Given that the typical stellar ages of these young quiescent galaxies are

Table 3
References Used in Figure 6 (in Descending Order of Observed Redshifts)

| (1) Data Reference | (2) N | (3) $\log(M_\star/M_\odot)$ | (4) $z_{obs}$ | (5) Quiescent Criterion | (6) Type | (7) Type Selection |
|---|---|---|---|---|---|---|
| This work (Blue Jay) | 14 | 10.3–11.7 | 1.7–2.7 | 1 dex below MS | RQ | $t_q^{lbt} < 0.5$[a] |
| ... | ... | ... | ... | ... | PSB | $t_q^{lbt} < 0.5$ and $t_{form}^{lbt} < 1.0$ |
| D. J. Setton et al. (2023) | 2753 | >11.2 | 0.4–0.8 | 0.6 dex below MS | RQ | $f_{1\,Gyr} > 0.1$[b] |
| ... | ... | ... | ... | ... | PSB | $f_{1\,Gyr} > 0.2$ |
| P.-F. Wu et al. (2018) | 467 | >10.0 | 0.4–0.8 | UVJ | RQ | EW(H$\delta$) > 4 Å |
| S. Tacchella et al. (2022a) | 105 | >10.0 | 0.6–1.0 | sSFR < $0.05/t_{univ}$ | RQ | $t_q^{lbt} < 1.0$ and $\tau_{quench} < 1.0$ |
| V. Wild et al. (2016) | 207 | >10.0 | 0.5–2.0 | Super-color | RQ | Super-color |
| K. E. Whitaker et al. (2012a) | 1519 | >10.7 | 1.2–2.2 | UVJ | RQ | $V − J < 0.9$ |
| S. Belli et al. (2019) | 5335 | >10.8 | 1.0–2.5 | UVJ | PSB | $0.3 < t_{50}/Myr < 0.8$[c] |
| C. D'Eugenio et al. (2020) | 9[d] | 10.8–11.5 | 2.4–3.3 | UVJ | PSB | $t_{50}/Myr < 0.8$ |
| B. Forrest et al. (2020) | 8[e] | >11.0 | 2.8–3.5 | 1 dex below MS | RQ | $t_q^{lbt} < 1.0$ |
| ... | ... | ... | ... | ... | PSB | $t_q^{lbt} < 1.0$ and $t_{form}^{lbt} < 1.0$ |
| C. Schreiber et al. (2018) | 21[f] | >10.3 | 3.0–4.0 | UVJ | RQ | $t_q^{lbt} < 1.0$ |
| ... | ... | ... | ... | ... | PSB | $t_q^{lbt} < 1.0$ and $t_{form}^{lbt} < 1.0$ |
| T. Nanayakkara et al. (2024) | 6 | 10.2–11.5 | 3.0–4.0 | UVJ | RQ | $t_q^{lbt} < 0.5$ |
| ... | ... | ... | ... | ... | PSB | $t_q^{lbt} < 0.5$ and $t_{form}^{lbt} < 1.0$ |
| A. C. Carnall et al. (2023b) | 10[g] | 10.1–11.4 | 3.0–5.0 | sSFR < $0.2/t_{univ}$ | PSB | SFH shapes |

**Notes.** Column (1): reference. Column (2): number of the total quiescent sample. Column (3): mass range of the quiescent sample. Column (4): observed redshift. Column (5): selection criteria for the quiescent sample. Column (6): type of quiescent galaxy for which the fraction was measured: RQs or PSBs. Column (7): selection criteria used for the type of quiescent galaxy.

[a] $t_q^{lbt}$: the quenching epoch in lookback time in Gyr.
[b] $f_{1Gyr}$: the fraction of stellar mass formed within the last Gyr.
[c] $t_{50}$: median stellar ages derived based on UVJ colors.
[d] Their stacked spectrum of nine quiescent galaxies at $2.4 < z < 3.3$ shows a strong Balmer break and strong H$\delta$ (EW > 10 Å), indicating PSBs.
[e] From their sample of 24 massive galaxies, we selected 8 galaxies that have sSFR 1 dex below SFMS.
[f] This includes 12 galaxies with $z_{spec}$ and 9 galaxies without $z_{spec}$.
[g] Robust sample.

~1 Gyr (e.g., K. E. Whitaker et al. 2012b), this selection would identify both PSBs and just RQs. Indeed, all our quiescent galaxies in Group-2 + Group-3 (except for COSMOS-11494) also satisfy H$\delta > 5$ Å. Therefore, we compare the fraction of young quiescent galaxies with that of our rapidly quenched population in panel (a).

We compare our PSBs with young quiescent galaxies in the literature when their reconstructed SFHs directly show recent starbursts or when they are as young as our PSBs (~0.5 Gyr). For example, D. J. Setton et al. (2023) derived SFHs from SED fitting, calculated the fraction of stellar mass formed within the last Gyr ($f_{0.1\,Gyr}$), and identified recently quenched galaxies having $f_{0.1\,Gyr} > 0.1$. Also, S. Belli et al. (2019) used the rest-frame UVJ colors to select the youngest quiescent galaxies with inferred stellar ages of 300–800 Myr. Therefore, we compare the results from these two studies with our PSB fraction. For the studies observing quiescent galaxies at high redshifts ($z > 3$), they are not representative samples and have small sample sizes. Based on the SFHs of individual quiescent galaxies presented in each work, we classify them into different groups and measure the fraction of rapidly quenched (Group-2 + Group-3) and PSBs (Group-3). We summarize all references in Table 3: how they selected their quiescent sample and RQs (or PSBs). In Figure 6, PSBs are/PSBs are selected using spectroscopic data (or inferred SFHs) as filled markers and using photometric data/colors as empty markers.

Approximately, 60% (9/14) of our Blue Jay quiescent galaxies are RQs, and only 30% (4/14) are rapidly quenched after "major" starbursts (PSBs). Both the RQ and PSB fractions are in good agreement with other studies and follow the trend of increasing fractions with redshift. At lower redshifts ($z < 1$), both RQs and PSBs are generally rare (only a few percent to <10%). On the other hand, in theory, all massive quiescent galaxies found at higher redshifts ($z > 4$) should be PSBs, accounting for their high-stellar-mass buildup and quenching within the age of the Universe. At $z = 2$–3, there is the largest difference between the RQs and PSBs, due to the population of galaxies in Group-2.

The galaxies in Group-2—rapidly quenched without major starbursts—appear to be the most commonly found at $z \sim 2$. It is unclear whether they truly had steady formation until recent quenching or had repeated bursts in the past. One possibility is that they might be the descendants of PSBs quenched at $z > 3$; as mergers are the most frequent at $z \sim 2$, PSBs quenched at $z > 3$ might have been rejuvenated with mergers and quenched again at $z \sim 2$. In this scenario, their reconstructed SFHs would show repeated bursts and quenching. These galaxies are still young quiescent galaxies, but they are slightly older than the PSBs, as they have formed the bulk of their stellar mass during the past burst events.

Note that in Figure 6, we include only the galaxies that are quenched "recently." We do not include galaxies rapidly quenched in the past, as we cannot constrain well the quenching speed of past quenching events. Thus, the fraction of recent RQs shown in panel (a) would be the lower limit of the fraction of all RQs. The fraction of all RQs would be much higher at lower redshifts. For example, S. Tacchella et al. (2022a) directly measured the quenching timescales and roughly half of their sample is rapidly quenched (with quenching timescales <1 Gyr). Also, Y. Zheng et al. (2022) found that ≈60% of quiescent galaxies at $z \sim 1$ in the SIMBA simulation are rapidly quenched.

## 4. Neutral and Ionized Gas

In this section, we study the neutral and ionized gas properties of our massive quiescent sample to understand the quenching mechanisms. We use the Na I D $\lambda5891$, $\lambda5897$ doublet as a tracer for the neutral gas, and study [N II] and H$\alpha$ emission for the ionized gas. We direct readers to R. L. Davies et al. (2024) for the full census of Na I D absorption at $z \sim 2$ from the representative Blue Jay sample. They found that half of the massive galaxies ($\log(M_\star/M_\odot) > 10$) show Na I D absorption and both star-forming and quiescent systems show similar incidences. For ionized gas, we only study [N II]$\lambda6550$, $\lambda6585$ and H$\alpha$ emission. See L. Bugiani et al. (2024) for the detailed study of all ionized emission lines (0.35–1.2 μm in rest frame) found in the quiescent galaxies from the Blue Jay survey.

Each panel in Figure 7 shows the Na I D absorption (left) and H$\alpha$ and [N II]$\lambda6550$, $\lambda6585$ emission (right). Galaxies are grouped in the same way as in Figure 4 into Group-1, Group-2, and Group-3, based on their formation and quenching history. The observed spectrum divided by the best-fit stellar model is shown for the Na I D absorption, and the observed spectrum subtracted from the stellar model is shown for H$\alpha$ and [N II] emission. A total of 12 of the 14 massive quiescent galaxies show Na I D excess by neutral gas. COSMOS-21477 does not have Na I D detection, and there are no spectrum data around Na I D for COSMOS-8469 (it falls in a detector gap). The panels for Na I D absorption excess are colored by whether the Na I D line is blueshifted (colored as sky blue) or consistent with galaxy systemic velocity (colored as yellow–green), based on the measurements presented in R. L. Davies et al. (2024; see their Table 1 for the detailed measurements of the centroid velocity offset, $\Delta v$). We fit the emission-line complex with three Gaussian components, each accounting for the H$\alpha$ and [N II]$\lambda6550$, $\lambda6585$ lines. The three lines share the same width, $\sigma_{gas}$, and we fix the flux ratio between [N II]$\lambda6550/\lambda6585$ to be 0.326, following N. M. Förster Schreiber et al. (2019). The [N II]/H$\alpha$ ratios are shown in the upper right corner of each panel.

We find a clear difference in the neutral gas properties between galaxies quenched early versus recently: in all early-quenched old galaxies (Group-1), the centroid of the Na I D absorption is consistent with the galaxy systemic velocity; in contrast, 75%[16] of the recently quenched galaxies (Group-2 and Group-3) show blueshifted Na I D excess absorption (more than 100 km s$^{-1}$), suggesting neutral gas outflows. The excess Na I D absorption is unlikely due to the enhanced sodium abundances of the stellar population; as tested in R. L. Davies et al. (2024), the excess absorption by extremely Na-enhanced stellar populations (e.g., [Na/Fe] = +0.6) is still weaker than the observed excess absorption in our sample. We interpret the neutral gas at systemic velocity as being due to either the cold-gas reservoir and/or slow outflows that cannot be resolved by the medium-resolution ($R \sim 1000$) gratings. On the other hand, the blueshifted absorption is a clear sign of outflowing gas.

Thus, we conclude that neutral gas outflows are more common and/or more powerful in galaxies that were recently quenched than in older galaxies.

This trend is consistent with the results from other studies based on local low-SFR galaxies. For example, Y. Sun et al. (2024) have measured the velocity offsets of Na I D excess absorption in local massive galaxies and found that the neutral gas outflow incidence and speed decrease with stellar age. They further selected a PSB sample (by requiring low H$\alpha$ emission and strong H$\delta$) and found that the outflow speed decreases with time elapsed since the starburst event.

The neutral gas outflows in the recently quenched galaxies are most likely driven by ejective AGN feedback. In R. L. Davies et al. (2024), they computed the energy and momentum rates of the neutral gas outflows and compared them to luminosity and momentum flux from young stars. They concluded that the star formation activity is insufficient to power the neutral outflows in the low-SFR galaxies. The right panels of Figure 7 show that some of the RQs (e.g., COSMOS-10565, 18668, 11142, and 8013) also show ionized H$\alpha$ and [N II] emission, fitted with high [N II]/H$\alpha$ ratios (shown in each panel). The high [N II]/H$\alpha$ ratios of the RQs seem to be more consistent with the AGN diagnostics (e.g., L. J. Kewley et al. 2001; G. Kauffmann et al. 2003), consistent with the results from the previous studies of S. Belli et al. (2024), R. L. Davies et al. (2024), and L. Bugiani et al. (2024). Therefore, the line ratio of ionized gas emission also supports that outflowing gas is most likely driven by AGN activities.

However, given the kinematics of ionized and neutral gas outflows, the bulk of outflowing materials will most likely remain in the halo. We measure the kinematics of ionized gas from the Gaussian fitting and find that the total line velocity dispersion is $\sigma_{gas} \sim 200$–600 km s$^{-1}$ (corresponding FWHM $\sim$ 470–1410 km s$^{-1}$). The outflowing velocities of the neutral gas in the RQs are estimated to be 200–700 km s$^{-1}$ (see R. L. Davies et al. 2024 for the detailed measurements), with large uncertainties as the Na doublets are not resolved with $R \sim 1000$ resolution. The kinematics of ionized and neutral gas outflows seem lower than the expected halo escape velocities, suggesting that the outflowing materials would still be gravitationally bound in the system (e.g., A. Concas et al. 2019). The gas remaining in the halos could potentially be reused to rejuvenate the galaxies. Other gas-heating processes, such as AGN radio-mode feedback (e.g., D. J. Croton et al. 2006; R. S. Somerville et al. 2008) or virial shock heating (e.g., S. D. M. White & M. J. Rees 1978; D. Kereš et al. 2005), are required to keep galaxies quenched over long timescales.

## 5. Number Density

We explore the number density evolution of quiescent galaxies using our representative mass-selected sample at $1.7 < z < 3.5$. We assume the area ($A$) the Blue Jay survey covered as the size of the two NIRSpec masks: $A = 2 \times 3.'6 \times 3.'4 = 0.0068$ deg$^2$. Due to the configuration of the MSA, the Blue Jay massive sample (74 galaxies with $\log(M_{star}/M_\odot) > 10$) does not include all the massive galaxies within the two mask regions. To assess the degree of spectroscopic completeness, we divide the total number of massive galaxies that lie within the footprint of the two masks by the number of massive galaxies for which we obtain a spectrum, obtaining a "scaling factor" of ≈3. However, the region probed by our observations is biased, because the two Blue Jay pointings were designed to maximize the number of massive targets and are therefore likely to probe overdense regions. This is confirmed by the presence of a known protocluster at $z \sim 2.1$ (T. Yuan et al. 2014) in the region observed by the Blue Jay survey. For this reason, we derive the scaling factor using the number density of massive galaxies in the entire CANDELS/COSMOS field, using the 3D-HST catalog, obtaining a factor of 2.35. Our estimated number density is multiplied by this scaling factor.

---

[16] Among the eight RQs in Group-2 and Group-3 (COSMOS-8469 in Group-2 has not been taken into account, as the region around Na D for this galaxy falls into the detector gap), six of them show blueshifted Na I D excess absorption.

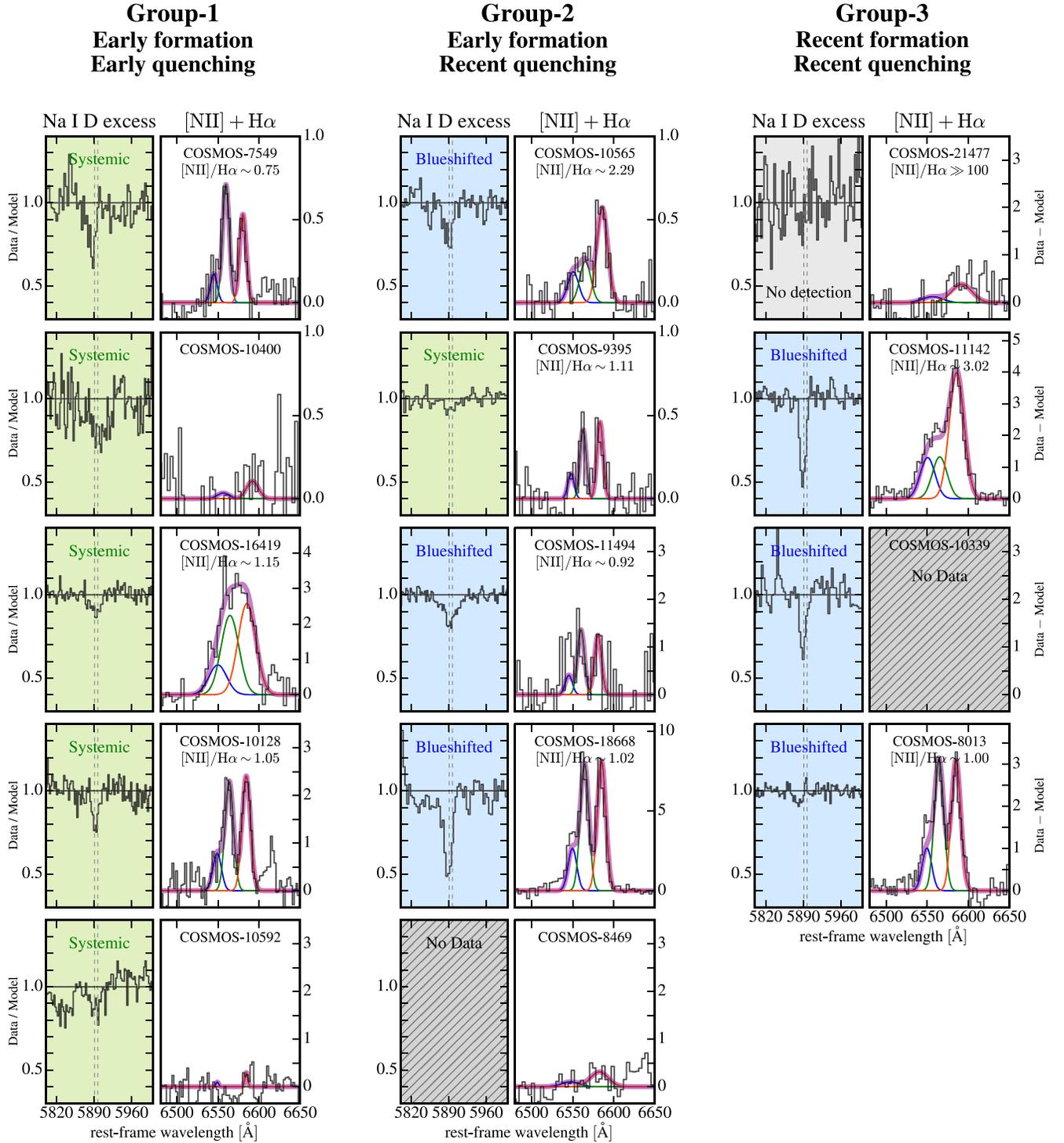

**Figure 7.** Na I D absorption by neutral gas (left) and Hα and [N II]λ6550, λ6585 emission (right) of the 14 massive quiescent galaxies. The observed data divided by the best-fit stellar model are plotted for Na I D absorption, and for ionized gas emission the observed data subtracted by the best-fit model are shown. The background of each Na I D panel is color-coded by the offset of centroid velocity of the Na I D line (green: same as the galaxy systemic velocity; blue: blueshifted by at least ~100 km s$^{-1}$). Three Gaussian components are fitted for the ionized gas emission complex, each accounting for Hα and [N II]λ6550, λ6585 emission. We find that almost all massive quiescent galaxies (except for COSMOS-21477) show neutral gas probed by Na I D absorption. Most of the recently quenched galaxies (Group-2 and Group-3) show neutral gas outflows (blueshifted Na I D absorption) and ionized gas emissions with high [N II]/Hα ratios. This supports the scenario of AGN activities driving multiphase outflows, eventually leading to recent rapid quenching.

The scaled number density of quiescent galaxies from our Blue Jay sample at $1.7 < z < 3.5$ is shown as a filled green star in Figure 8: $n_{\rm qsc} = 2.72 \pm 0.64 \times 10^{-4}$ Mpc$^{-3}$. The error bar in redshift covers the entire redshift range of our sample ($z = 1.7$–$3.5$), and the error bar in number density is from the Poisson uncertainty. Note that the number density we derived for $1.7 < z < 3.5$ is based on the assumption that the fractions of quiescent galaxies in the protocluster and field regions would be similar. However, some studies have reported that the quiescent fractions are higher in the overdense/protocluster regions (e.g., I. McConachie et al. 2022; K. Ito et al. 2023; M. Tanaka et al. 2024). Given that our Blue Jay galaxies are observed in the protocluster region, there is a possibility that the number density of massive quiescent galaxies we derived is higher than that of the average field at $z \sim 2$.

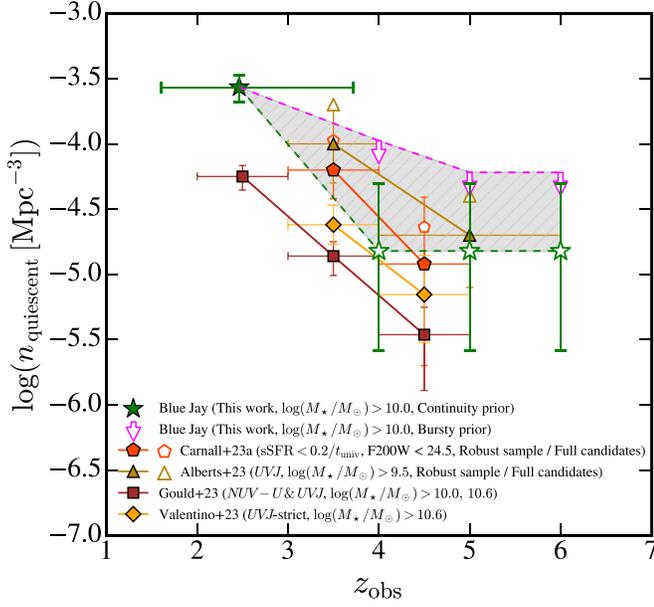

**Figure 8.** Number density of massive quiescent galaxies as a function of redshift. The filled green star is the number density of the Blue Jay quiescent galaxies at $z \sim 2$. We predict the number densities of massive quiescent galaxies at higher redshifts ($z = 4$–$6$) by tracing back the SFHs of our entire representative sample using the continuity prior, shown as empty green stars. We also show the predicted number densities based on the SFHs derived using the bursty prior as magenta arrows, and these estimates are likely the upper bounds. The true number densities at $z > 4$ would likely fall between the two estimates: $n_{\rm qsc} \approx (1.5$–$6.0) \times 10^{-5}$ Mpc$^{-3}$ ($z = 4$–$6$), shown as the gray hatched region. The number densities of massive quiescent galaxies directly measured by other recent studies are also shown as different symbols: A. C. Carnall et al. (2023b; orange pentagons), K. M. L. Gould et al. (2023; brown squares), F. Valentino et al. (2023; yellow diamonds), and S. Alberts et al. (2024; gold triangles). We find that our predicted number densities at $z > 3$ are broadly consistent with the number densities measured at high redshifts.

We further estimate the number density of quiescent galaxies at $z = 4, 5$, and $6$ by tracing back the SFHs of all our 141 representative sample galaxies. We identify quiescent galaxies as the galaxies having sSFR $< 0.2/t_{\rm univ}(z)$ (we take the median SFRs of the posteriors). In Figure 8, the number densities at $z > 3$ are shown as empty green stars: $n_{\rm qsc,continuity} = 1.51^{+3.47}_{-1.25} \times 10^{-5}$ Mpc$^{-3}$ ($z = 4$–$6$).[17] The predicted number densities from the Blue Jay survey are also multiplied by the scaling factor. The error bars and shaded region include the Poisson errors. These estimated number densities at $z = 4$–$6$ are based on one massive quiescent galaxy already quenched at $z \gtrsim 6$ (COSMOS-16419). None of the star-forming galaxies in our Blue Jay sample have been quenched before the observed epochs, which might be due to the continuity prior used to derive SFHs, as it disfavors abrupt changes in SFH, including rejuvenation.

---
[17] The $1\sigma$ confidence level for small-number Poisson statistics is calculated following N. Gehrels (1986).

We also calculate the number density from the SFHs derived using the bursty prior. When the bursty prior is used, galaxies show quenching and rejuvenation more frequently throughout their history, increasing the predicted number density of quiescent galaxies at high redshifts. However, the increase is not dramatic, since we only consider massive galaxies ($\log(M_\star/M_\odot) > 10$) at each redshift. If galaxies formed most of their mass during recent rejuvenation, they would be below the mass cut at high redshifts. Thus, even if they are quiescent at high redshifts, they would not contribute to the number density estimates for massive quiescent galaxies. We remove galaxies where we fit only the observed photometry (see Section 2.2) because, without spectral features to fit, SFH is not very well constrained with the bursty prior. The number densities derived using the bursty prior are likely the upper bounds, so we show them as empty magenta arrows in Figure 8: $n_{\rm qsc,bursty} = 1.06 \pm 0.26 \times 10^{-4}$ Mpc$^{-3}$ ($z = 4$) and $n_{\rm qsc,bursty} = 6.04 \pm 1.97 \times 10^{-5}$ Mpc$^{-3}$ ($z = 5, 6$). Considering the caveats of both the continuity and the bursty priors, the true number densities at $z > 4$ would likely fall between the two estimates: $n_{\rm qsc} \approx (1.5$–$6.0) \times 10^{-5}$ Mpc$^{-3}$ ($z = 4$–$6$), shown as the gray hatched region in Figure 8.

We compare the predicted number densities with those measured from recent observations of massive quiescent galaxies at $z > 3$. For example, A. C. Carnall et al. (2023b) used the sSFR cut (sSFR $< 0.2/t_{\rm univ}(z)$) to select quiescent galaxies from the CEERS survey. Their number density estimates for the quiescent galaxies with $F200W < 24.5$ (equivalent to $\log(M_\star/M_\odot) \gtrsim 10.5$) are shown as orange pentagons in Figure 8 (filled: robust sample; empty: full candidates including galaxies with secondary star-forming solutions). K. M. L. Gould et al. (2023) adopted a Gaussian Mixture Model (GMM) trained on rest-frame colors ($NUV - U$, $U - V$, and $V - J$) to select massive quiescent galaxies ($\log(M_\star/M_\odot) > 10.6$) from the COSMOS2020 survey. The number densities found in their study are shown as brown rectangles. F. Valentino et al. (2023) also identified quiescent galaxies from the public JWST fields using a similar approach, with a GMM trained with rest-frame colors, and measured the number density, shown as orange diamonds. S. Alberts et al. (2024) found the number densities of quiescent galaxies with $\log(M_\star/M_\odot) > 9.5$ from their robust and full $UVJ$-selected quiescent galaxies from JADES (shown as filled and empty gold triangles). All the error bars in redshift in Figure 8 cover the entire redshift range of galaxies at each redshift bin.

Broadly, our predicted number densities of massive quiescent galaxies at $z > 3$ are consistent with the results of A. C. Carnall et al. (2023b) and S. Alberts et al. (2024). Our number density estimates derived using the bursty prior (magenta arrows) are slightly higher. However, these estimates are likely upper bounds and still within the uncertainty of the number densities derived using the full quiescent candidates in both studies (empty markers). Our predicted number density is still higher than what is found in K. M. L. Gould et al. (2023) and F. Valentino et al. (2023), which might be due to the different stellar mass ranges of the quiescent sample and more stringent criteria they used for quiescent selection. The galaxies in both studies are slightly more massive ($\log(M_\star/M_\odot) > 10.6$) than ours ($\log(M_\star/M_\odot) > 10.0$). Both studies have used a GMM trained on rest-frame colors to select quiescent galaxies. However, two of our 14 quiescent galaxies (COSMOS-8469 and 18668) are located near the boundaries of the rest-frame $UVJ$ cut for the quiescent region (see Figure A1). Moreover, the comparison with other works is affected by cosmic variance, which is particularly important for small fields and massive galaxies (C. L. Steinhardt et al. 2021; F. Valentino et al. 2023). Finally, we point out that our result is based on a spectroscopic sample of galaxies that lie in an overdense region. While we correct for this effect in terms of the overall number density, as explained above, it is still possible that the inferred SFHs for our sample are biased compared to those of the galaxy populations living in other types of environment.

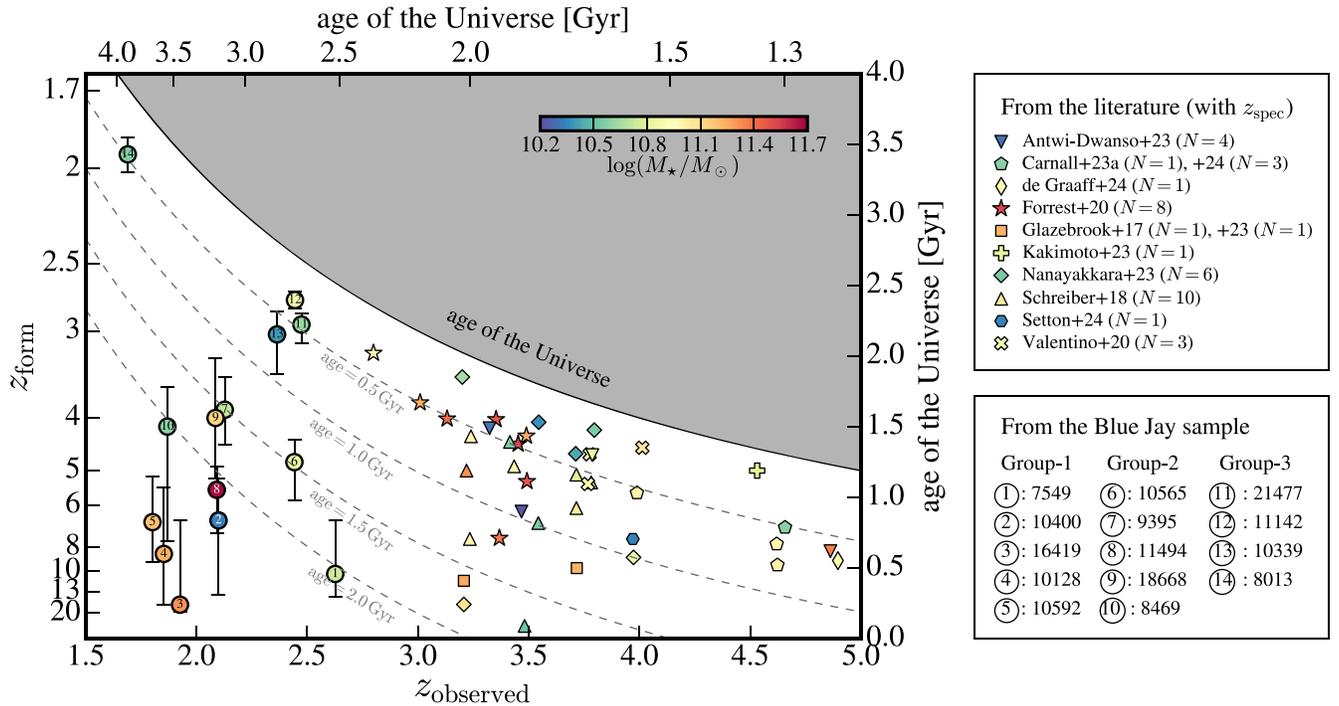

**Figure 9.** The formation redshift ($z_{form}$) shown as a function of the observed redshift ($z_{observed}$). Galaxies are color-coded by their stellar mass. The dashed curves indicate lookback times of 0.5, 1.0, 1.5, and 2.0 Gyr (or the "age" of a galaxy). The 14 Blue Jay massive quiescent galaxies are shown as circles with their IDs. Massive quiescent galaxies found at $z > 3$ collected from the literature are also plotted with different symbols.

## 6. Discussion

### 6.1. Descendants of Early (z > 3) Quiescent Galaxies at z ~ 2

In Figure 9, we collect all the spectroscopically confirmed massive quiescent galaxies at $z_{spec} > 3$ along with the Blue Jay quiescent galaxies and put them in the plane of formation redshift ($z_{form}$) and observed redshift ($z_{obs}$). The massive quiescent galaxies from the literature are summarized in the right panel (K. Glazebrook et al. 2017, 2024; C. Schreiber et al. 2018; B. Forrest et al. 2020; F. Valentino et al. 2020; T. Nanayakkara et al. 2024; J. Antwi-Danso et al. 2023; A. C. Carnall et al. 2023a, 2024; T. Kakimoto et al. 2024; A. de Graaff et al. 2024; D. J. Setton et al. 2024), with the number of samples ($N$) identified in each study. The Blue Jay galaxies are shown as circles with their IDs. All galaxies are color-coded by their stellar mass. The black curve represents the age of the Universe. The dashed curves show lookback times of $t_{lbt} = 0.5$, 1.0, 1.5, and 2.0 Gyr, so galaxies on these curves would have stellar ages of roughly ~0.5, 1.0, 1.5, and 2.0 Gyr. Massive quiescent galaxies at early epochs ($z > 3$) in the literature have a wide range of formation redshifts, from $z_{form} \approx 3$ (recent formation) to $z_{form} > 6$ (early formation).

Several studies have confirmed that the majority of massive quiescent galaxies observed in the young Universe ($z > 3$) appear to have a PSB nature (e.g., C. D'Eugenio et al. 2020). As shown in Figure 9, indeed, the bulk of the massive quiescent galaxies observed at $z > 3$ are young populations (with stellar ages of ~0.5 Gyr), similar to our Blue Jay galaxies in Group-3 (most likely PSBs). If these early PSBs remain quiescent since quenching, they would be observed as past PSBs at $z \sim 2$. The number density of massive quiescent galaxies increases roughly by a factor of 10 from $z \sim 3.5$ to $z \sim 2$ (see, e.g., the results of K. M. L. Gould et al. 2023 in Figure 8). Given that we have 14 massive quiescent galaxies, we should be able to see at least one or two descendants of early quiescent galaxies in our "early-formed" quiescent galaxies (Group-1 and Group-2).

In Group-2, we do not see any clear cases of past PSBs, even when using the bursty SFH prior (see the results in Figure 5) that can reproduce the past burst events (see Appendix D and Figure A6). However, we cannot rule out the possibility that some of the galaxies in Group-2 might be the descendants of the PSBs at $z > 3$. For example, in the case of COSMOS-9395, 10565, and 8469, based on the inferred SFHs using the bursty prior (see Figure 5), they appear to have a starburst ~1.0–1.5 Gyr ago followed by abrupt quenching. They then experienced another starburst and rapid quenching very recently, but since they formed the bulk of their mass at their past burst events ($z_{form} \sim 3–4$), they are older than PSBs formed with a single recent burst. However, it is unclear whether they truly had repeated bursts and quenching or rather had a steady formation with nearly constant SFRs (as shown by the continuity prior).

The galaxies in Group-1 are more likely candidates for the descendants of early PSBs. None of them have inferred SFHs similar to those of PSBs quenched in the past. However, since they are quenched at early epochs, the accretion of smaller galaxies (i.e., minor mergers) after quenching could have smoothed out their SFHs. Recently, from deep JWST imaging, K. A. Suess et al. (2023) have detected a large population of low-mass companions (down to <1:10 mass ratio) around massive quiescent galaxies at $0.5 < z < 3.0$. They found that most companions are bluer than their primary galaxy, therefore they would most likely comprise younger stellar populations. The accretion of these young stars could dilute the rapid quenching signal in the past, making it difficult to reconstruct this abrupt quenching event in the inferred SFHs of past PSBs.

Two of the old quiescent galaxies formed at very early epochs (COSMOS-7549 and COSMOS-16419) could be good candidates for high-$z$ PSB descendants. The estimated formation redshifts for both galaxies ($z_{form} \gtrsim 10$) are much higher than those of early PSBs (see Figure 9). However, we should be careful interpreting the estimated formation redshift of these old galaxies (with ages older than $\gtrsim 2$ Gyr), as it is very challenging to infer the exact ages of old stellar populations. For example, while the bursty prior can recover the presence of the past starburst event, the exact moment (e.g., the star formation peak) is still not very well reproduced if it is more than 2 Gyr ago. This is confirmed by our tests (see Appendix D for details): in Figure A6, the inferred star formation peak deviates by $\sim 0.5$ Gyr from the true star formation peak at $\gtrsim 2$ Gyr ago, which can be as much of a difference as between $z = 6$ and $z = 10$. Therefore, considering this uncertainty in inferring the early formation epochs, it is more plausible that they are formed at $z_{form} \gtrsim 6$. Nonetheless, it still holds true that these are formed and quenched at early times, making good candidates for high-$z$ PSB descendants. Here we discuss these two galaxies in more detail.

*COSMOS-7549.* This galaxy is observed at the highest redshifts in our quiescent sample, at $z = 2.63$. The Prospector fit is shown in Figure A7, including the results using both the continuity and bursty priors. The stellar mass is $M_\star \sim 6.0 \times 10^{10} M_\odot$, and it formed half of its mass at $z_{form} \gtrsim 10$ with high SFR $\approx 200 \, M_\odot \, \text{yr}^{-1}$. The quenching history after its formation depends on the SFH prior. When the continuity prior is used, this galaxy is relatively gradually quenched at $z_{quench} = 3.97$. On the other hand, the bursty prior shows that the galaxy is quenched abruptly at a much earlier epoch, at $z_{quench} = 9.70$. Considering that (1) this galaxy is formed at a very early epoch $z_{form} \gtrsim 10$ (confirmed by using both SFH priors) and (2) the continuity prior is designed to disfavor abrupt quenching, and indeed does not seem to well reproduce the past rapid quenching event (Figure A6), it is more likely that the galaxy is quenched at the early epoch as well, as inferred using the bursty prior. The galaxy shows the Na I D excess absorption by neutral gas, with the centroid velocity consistent with the systemic velocity of the galaxy, likely tracing the ISM in the neutral phase.

*COSMOS-16419.* This galaxy is the second most massive galaxy in our quiescent sample: $M_\star = 2.1 \times 10^{11} M_\odot$ observed at $z = 1.925$. The Prospector fitting result of this galaxy is shown in Figure A8. It formed most of its stellar mass at $z_{form} \gtrsim 10$ with extremely high SFR $\approx 1000 \, M_\odot \, \text{yr}^{-1}$, and only after $\sim 300$ Myr was it rapidly quenched at $z_{quench} = 7–13$. This extremely early formation and quenching is confirmed even when using the continuity prior, which disfavors rapid changes in SFH. After abrupt quenching at a very early epoch, the SFR significantly drops to $\sim 1 \, M_\odot \, \text{yr}^{-1}$ and this galaxy appears to have been rejuvenated recently with SFR $\sim 10 \, M_\odot \, \text{yr}^{-1}$ (shown by both the continuity and the bursty prior). The weak H$\alpha$ emission (Figure 7) could also be due to the recent rejuvenation. The color-composite image in Figure A8(c) shows two small companions, which could potentially have contributed to the recent rejuvenation. This galaxy also shows Na I D absorption by neutral gas, with the centroid velocity consistent with systemic velocity. The neutral gas might originate from the cold-gas reservoir in the galaxy. Another possibility is that the galaxy might be in the process of quenching the recently rejuvenated star formation, and a fraction of the neutral gas could be outflows as yet unresolved due to the spectral resolution.

### 6.2. Physical Mechanisms Behind the Widespread Rapid Quenching at z ∼ 2

Almost all massive quiescent galaxies at $z \sim 2$ in our sample appear to be quenched rapidly. In Section 3, we classified galaxies into three groups according to their SFHs and found that 9 of 14 galaxies are rapidly quenched at recent epochs (Group-2 and Group-3). We also identified five galaxies that have been quenched for more than 0.5 Gyr (Group-1). While their estimated quenching speeds are lower than those of recently quenched galaxies, this might be a biased result, due to difficulties in inferring the exact ages of old stellar populations; in other words, any sharp transitions in SFH at old age bins would not be well recovered. Also, as discussed in Section 6.1, the accretion of smaller galaxies after quenching, which tend to be bluer (e.g., K. A. Suess et al. 2023), could dilute the rapid quenching signal in the past. Because quenching is generally expected/required to be rapid at higher redshifts (given the young age of the Universe), it is more likely that these old galaxies were also quenched rapidly in the past, but it is very challenging to recover these rapid transitions in the inferred SFHs. Therefore, we conclude that our sample seems to lack clear candidates for the "slowly quenched" galaxies.

The ejective AGN feedback seems to drive the widespread rapid quenching of massive quiescent galaxies at $z \sim 2$. Recent studies of S. Belli et al. (2024) and F. D'Eugenio et al. (2024) have detected deep blueshifted Na I D absorption, indicating outflowing neutral gas, in PSBs at $z \sim 3$. They found that the neutral outflow rates are 2 orders of magnitude higher than the ionized outflow rates and exceed the current SFRs of the host galaxy, suggesting that the ejected gas is massive enough to quench galaxies. R. L. Davies et al. (2024) have found that strong neutral gas outflows are widespread in massive galaxies at $z \sim 2$ and are most likely driven by ejective AGN feedback. In Section 4, we have found a clear trend where the neutral gas outflows, probed by blueshifted Na I D absorption, are more common and powerful in recently quenched galaxies than in older galaxies. Many of these recently quenched galaxies also show ionized gas emission with high [N II]/H$\alpha$ ratios, more consistent with AGN diagnostics (see also L. Bugiani et al. 2024). This all supports a picture where the strong AGN activity drives multiphase gas outflows, leading to rapid quenching.

The old quiescent galaxies in Group-1 do not show neutral gas outflows at the moment of observations, as they are quenched (probably rapidly) farther back in time. Interestingly, all of the five galaxies in Group-1 show Na I D excess absorption by neutral gas (Figure 7), and the centroid velocity is consistent with the galaxy systemic velocity. This indicates that neutral gas may come from either cold-gas reservoirs and/or slow outflows that cannot be resolved by the medium-resolution ($R \sim 1000$) gratings. The presence of neutral gas reservoirs might suggest that the ejective AGN feedback that led to rapid quenching in the past has weakened over time. Indeed, the two old galaxies, COSMOS-7549 and 16419, show weak H$\alpha$ emission possibly due to residual star formation. However, the other three galaxies stay quenched and have very low SFRs at the observed epochs. One possibility is that preventative AGN feedback might prevent this gas from collapsing into forming stars (e.g., N. Roy et al. 2021).

For the galaxies recently and rapidly quenched after a major starburst (Group-3), the gas inflow toward central regions, triggering the starbursts, seems to be a key driver of the strong AGN feedback. Many studies have shown that gas-rich mergers can induce starbursts in the central regions (e.g., J. E. Barnes & L. E. Hernquist 1991; K. Bekki et al. 2005; G. F. Snyder et al. 2011), making galaxies compact. The gas inflow can also fuel the central SMBH (e.g., P. F. Hopkins et al. 2006), and the resulting strong AGN activity can directly eject the gas from a galaxy, leading to rapid quenching. Indeed, gas-rich major mergers seem to be more common at $z \sim 2$ (e.g., C. J. Conselice et al. 2008). However, several studies have also shown that other compaction processes without mergers—for example, violent disk instability or the accretion of misaligned gas streams (e.g., A. Dekel & A. Burkert 2014; A. Zolotov et al. 2015; S. Tacchella et al. 2016)—can also trigger central starbursts, especially at high redshifts. While it is unclear whether mergers are always involved, the mechanism driving the gas inflow toward central regions seems to be responsible for the strong AGN feedback causing rapid quenching.

The physical mechanism behind the recently quenched galaxies after a flat or bursty history (in Group-2) is unclear. Ejective AGN feedback may be responsible for the recent rapid quenching in these galaxies as well, given the observed outflowing neutral gas. However, it is not clear why AGN feedback suddenly becomes effective and rapidly quenches galaxies. The formation history of these galaxies depends on the assumed SFH prior (see Section 3.2): steady formation with nearly constant SFRs (continuity prior) versus bursty formation with repeated bursts and quenching (bursty prior). If they indeed experience repeated bursts and quenching, their formation could be explained by a similar scenario (gas inflow toward central regions fueling the SMBH) as the galaxies in Group-3. However, mergers might be a more likely scenario for the galaxies in Group-2. Many studies using cosmological galaxy simulations have shown that the fraction of ex situ stars in a galaxy with a stellar mass of $M_\star \sim 10^{11} M_\odot$ at $z \sim 2$ could be as high as $\sim 20\%$ of the total stellar mass (e.g., V. Rodriguez-Gomez et al. 2016; S. Tacchella et al. 2019). A large population of accreted old stars could create a past burst in the inferred SFHs or smooth it out to be a flat history.

## 7. Summary and Conclusion

In this work, we have studied the star formation and quenching history of 14 massive ($\log(M_\star/M_\odot) > 10.0$) quiescent galaxies at $z \sim 2$. Our 14 quiescent galaxies are a mass-selected representative sample from the Blue Jay survey, a JWST Cycle 1 program that obtained NIRSpec $R \sim 1000$ spectra for 141 galaxies with $9 < \log(M_\star/M_\odot) < 11.5$ at $1.7 < z < 3.5$. The well-defined selection function of the Blue Jay sample allows us to study the full census of the massive quiescent population with robust statistics. We used Prospector SED fitting models to fit both observed photometry and the NIRSpec data to infer their SFHs. Here we summarize our results.

1. We find that massive quiescent galaxies can be split into three categories, according to their SFHs with roughly equal numbers of galaxies (Figure 4): (1) relatively old galaxies quenched at early epochs; (2) galaxies that are rapidly and recently quenched after a flat or bursty formation history (depending on the assumed prior); and (3) galaxies that are rapidly and recently quenched after a major starburst.

2. Almost all massive quiescent galaxies in our sample show Na I D absorption by neutral gas. Most recently quenched galaxies show blueshifted Na I D absorption, suggesting neutral gas outflows (see Figure 7). Some of them also show ionized gas emission, with high line ratios of [N II]/H$\alpha$, consistent with AGN diagnostics. This supports the scenario of AGN activity driving multiphase gas outflows, leading to recent rapid quenching. However, given the kinematics of ionized gas and the outflowing neutral gas, the bulk of the outflowing gas will most likely remain in the halos.

3. We measure the number density of massive quiescent galaxies at $z \sim 2$ from the Blue Jay quiescent sample: $n_{\rm qsc} = 2.72 \pm 0.64 \times 10^{-4}\,{\rm Mpc}^{-3}$. We further derive the number density of massive quiescent galaxies at $z > 3$ by tracing back the SFHs of our entire representative sample: $n_{\rm qsc} \approx (1.5-6.0) \times 10^{-5}\,{\rm Mpc}^{-3}$ at $z = 4-6$ (see Figure 8). Our predicted number density at high redshifts agrees well with other studies based on the most recent JWST observations (e.g., A. C. Carnall et al. 2023b).

4. The two old massive quiescent galaxies in our sample, COSMOS-7549 ($\log(M_\star/M_\odot) = 10.8$ at $z = 2.63$) and COSMOS-16419 ($\log(M_\star/M_\odot) = 11.5$ at $z = 1.93$), appear to be formed at very early epochs, $z_{\rm form} \gtrsim 6$. These two galaxies are likely the descendants of early PSBs at $z > 3$. Both of them are quenched at early epochs yet still appear to have neutral gas reservoirs and weak H$\alpha$ emission. This is possibly because the ejective AGN feedback that led to rapid quenching in the past has weakened over time.

At cosmic noon ($z \sim 2$), galaxy growth peaks with active star formation and frequent mergers. The resulting supernovae and SMBH feedback processes are also thought to be the most active, efficiently regulating galaxy growth. At the massive end, a large fraction of galaxies are quiescent at this epoch. Based on our study of massive quiescent galaxies at $z \sim 2$, we conclude that rapid quenching by strong AGN feedback driving multiphase outflows is the dominant quenching process at cosmic noon. We find that most of the massive quiescent galaxies ($\approx 70\%$) are rapidly quenched recently ($<0.5$ Gyr), newly added to the quiescent population. This suggests that the massive quiescent population experiences rapid growth at $z \sim 2$, possibly driven by active SMBH feedback.

## Acknowledgments

The Blue Jay Survey is funded in part by STScI Grant JWST-GO-01810. S.B. and L.B. are supported by the ERC Starting Grant "Red Cardinal," GA 101076080. B.D.J. acknowledges support by JWST-GO-01810. R.L.D. is supported by the Australian Research Council through the Discovery Early Career Researcher Award (DECRA) Fellowship DE240100136 funded by the Australian Government. J.L. and Y.L. are supported under Program number JWST-GO-01810.004-A, provided through a grant from STScI under NASA contract NAS503127. R.E. acknowledges the support from grant Nos. 21-atp21-0077, NSF AST-1816420, and HST-GO-16173.001-A, as well as from the Institute for Theory and Computation at the Center for Astrophysics. R.W. acknowledges the funding of a Leibniz Junior Research Group (project number J131/2022).

This work is based on observations made with the NASA/ESA/CSA James Webb Space Telescope. The JWST data presented in this article were obtained from the Mikulski Archive for Space Telescopes at the Space Telescope Science Institute (and can be accessed via doi:10.17909/t11e-xh37), which is operated by the Association of Universities for Research in Astronomy, Inc., under NASA contract NAS 5-03127 for JWST. These observations are associated with program GO 1810. This work also makes use of observations taken by the 3D-HST Treasury Program (GO 12177 and 12328) with the NASA/ESA HST, which is operated by the Association of Universities for Research in Astronomy, Inc., under NASA contract NAS5-26555. The photometric catalog of all galaxies from the Blue Jay survey is publicly available at doi:10.5281/zenodo.13292819. The JWST spectra and the best-fit SED models for the quiescent sample studied in this work are presented at doi:10.5281/zenodo.13365382.

## Appendix A
## Quiescent Sample in the Rest-frame *UVJ* Diagram

Figure A1 shows the rest-frame *UVJ* color–color diagram of all our 141 Blue Jay galaxies. Galaxies are color-coded by their stellar ages, and the size of the circles indicates their stellar mass. As in Figure 2(a), galaxies for which only observed photometry is fitted are shown as crosses, and the circles show the galaxies where we fit both observed photometry and the NIRSpec data. The gray shaded region is the quiescent region defined in A. Muzzin et al. (2013). Most of the 14 massive quiescent galaxies identified using sSFR would also be selected as quiescent by the rest-frame *UVJ* colors. COSMOS-8469 and 18668 are outside the quiescent region but near its border. See L. Bugiani et al. (2024) for more details on the emission-line

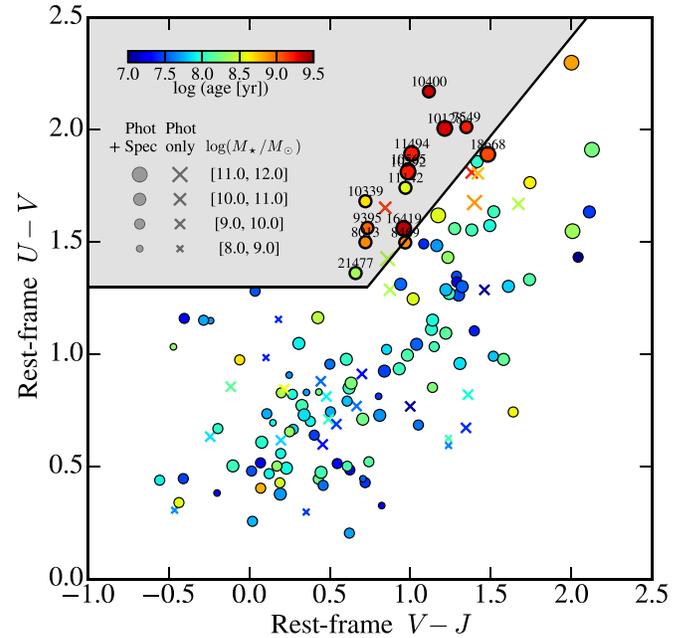

**Figure A1.** Rest-frame *UVJ* diagram. Galaxies are color-coded by their light-weighted ages derived from Prospector fitting. Most of the 14 massive quiescent galaxies selected by sSFR (1 dex below the SFMS) would also be identified as quiescent by the rest-frame *UVJ* criterion. Two of the 14 galaxies, COSMOS-8469 and 18668, are outside the *UVJ* quiescent region but near its border.

properties of galaxies as a function of their position on the *UVJ* diagram.

## Appendix B
## NIRSpec Data and Prospector Fitting

We present the NIRSpec $R \sim 1000$ data obtained from the Blue Jay survey and the best-fit SED models from Prospector fitting in Figures A2, A3, and A4. For each galaxy, panel (a) shows the photometric data we used with circles (blue: HST/ACS+WFC3; green: JWST/NIRCam), and the best-fit SED model is shown with the red line. The blue line in panel (b) shows the NIRspec spectrum (in arbitrary units) between 3700 and 6700 Å in the rest frame. Only 4000 and 6700 Å in the rest-frame wavelength range is used for fitting. The yellow–green shade shows the part of the spectrum used for fitting. Note that the Na I D line is masked out. Panel (c) shows the color-composite image of the galaxy in a $5'' \times 5''$ box. The image is an RGB color-composite image using F090W (blue), F200W (green), and F444W (red) bands from the PRIMER survey (GO: 1837; PI: J. Dunlop).

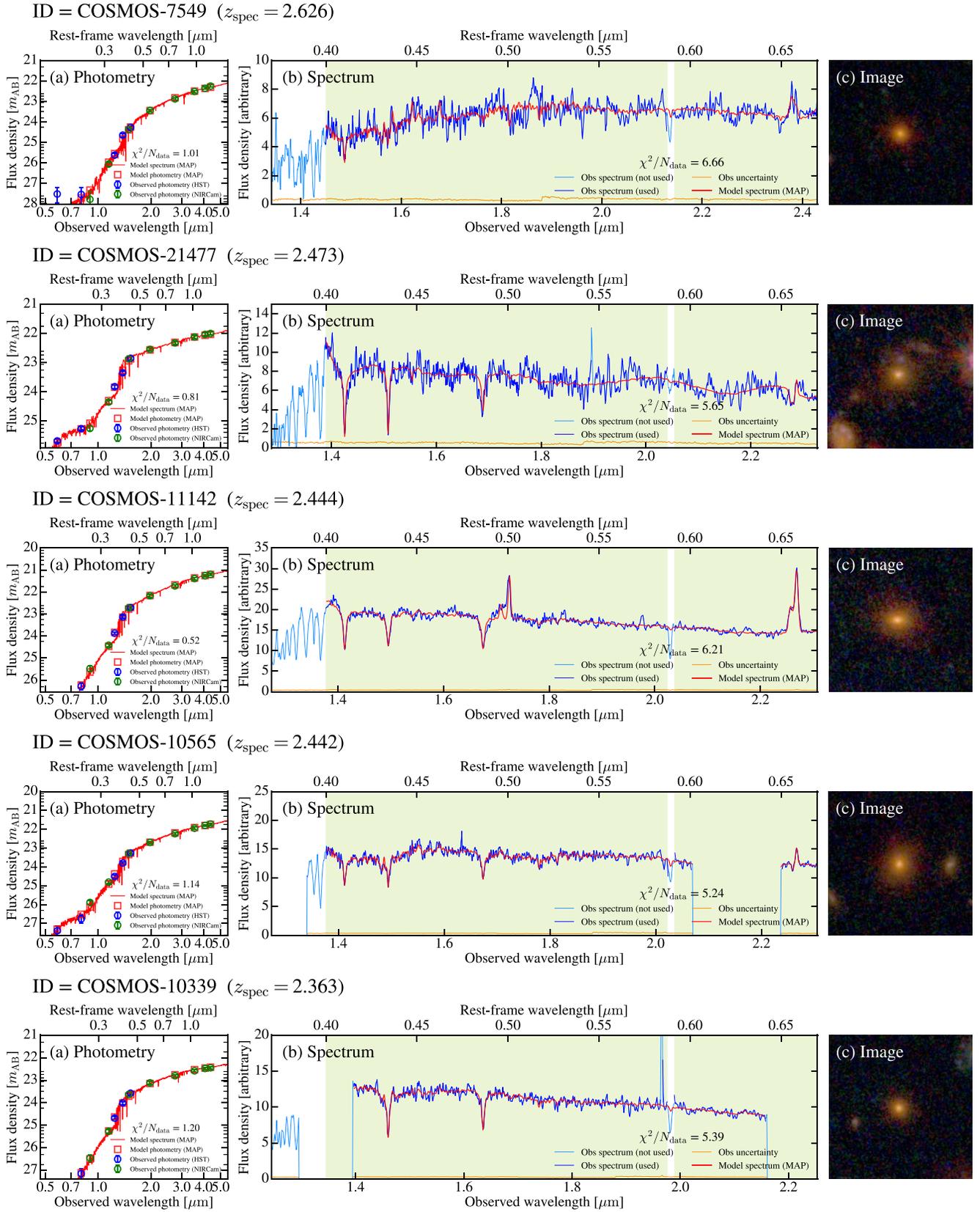

**Figure A2.** Best-fit SED fitting to both observed photometry and spectrum. (a) Observed HST and NIRCam photometry in blue and green circles, respectively. The best-fit SED model is shown with the red line. (b) NIRSpec spectrum between 0.37 and 0.67 $\mu$m in the rest frame, shown with the sky-blue line. The green shaded region indicates the wavelength range used for fitting, and the spectrum data used for fitting are highlighted as the blue line. Both the observed spectrum and best-fit model are smoothed by 3 pixels. (c) Color-composite image in $5'' \times 5''$ box.

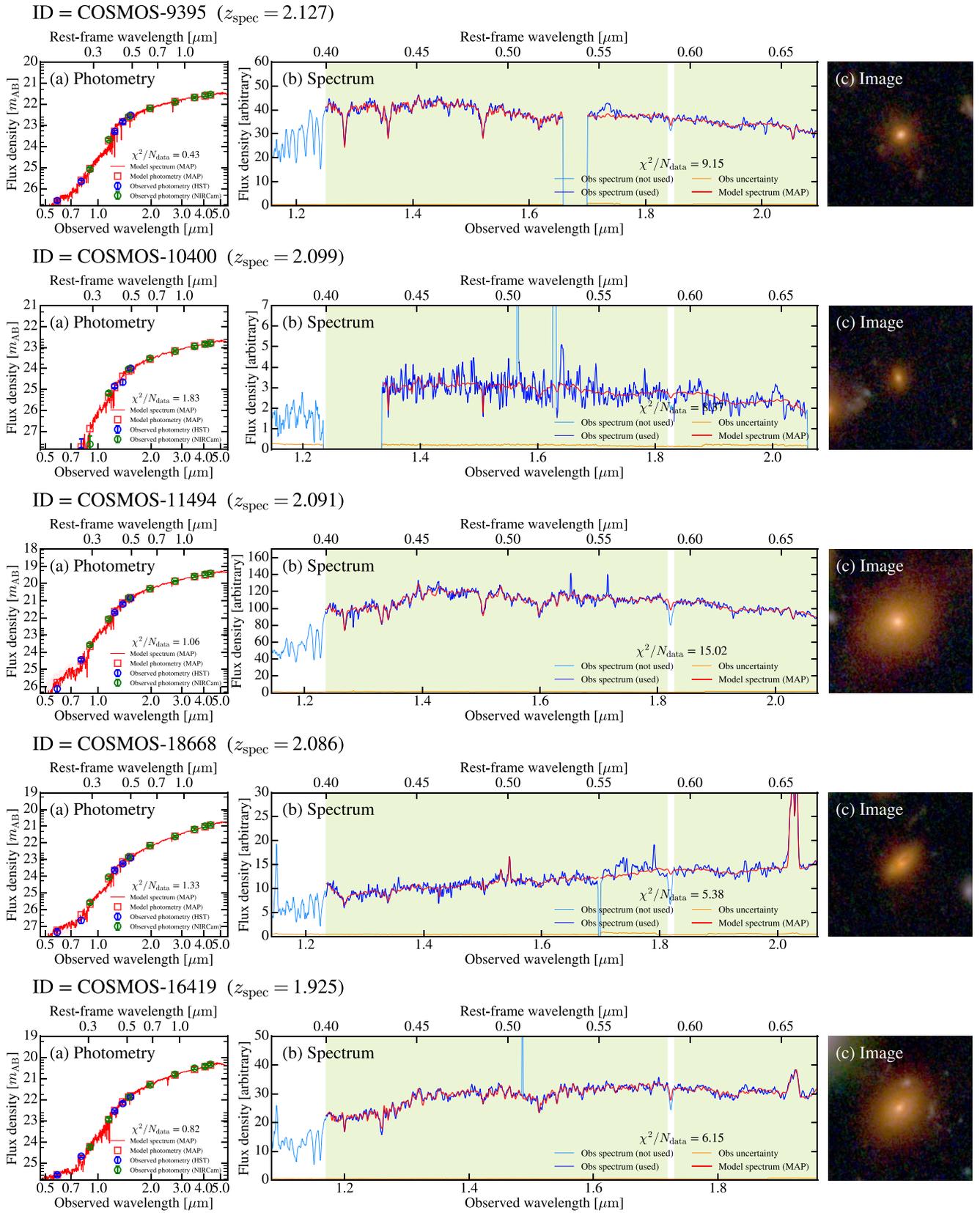

**Figure A3.** Continued.

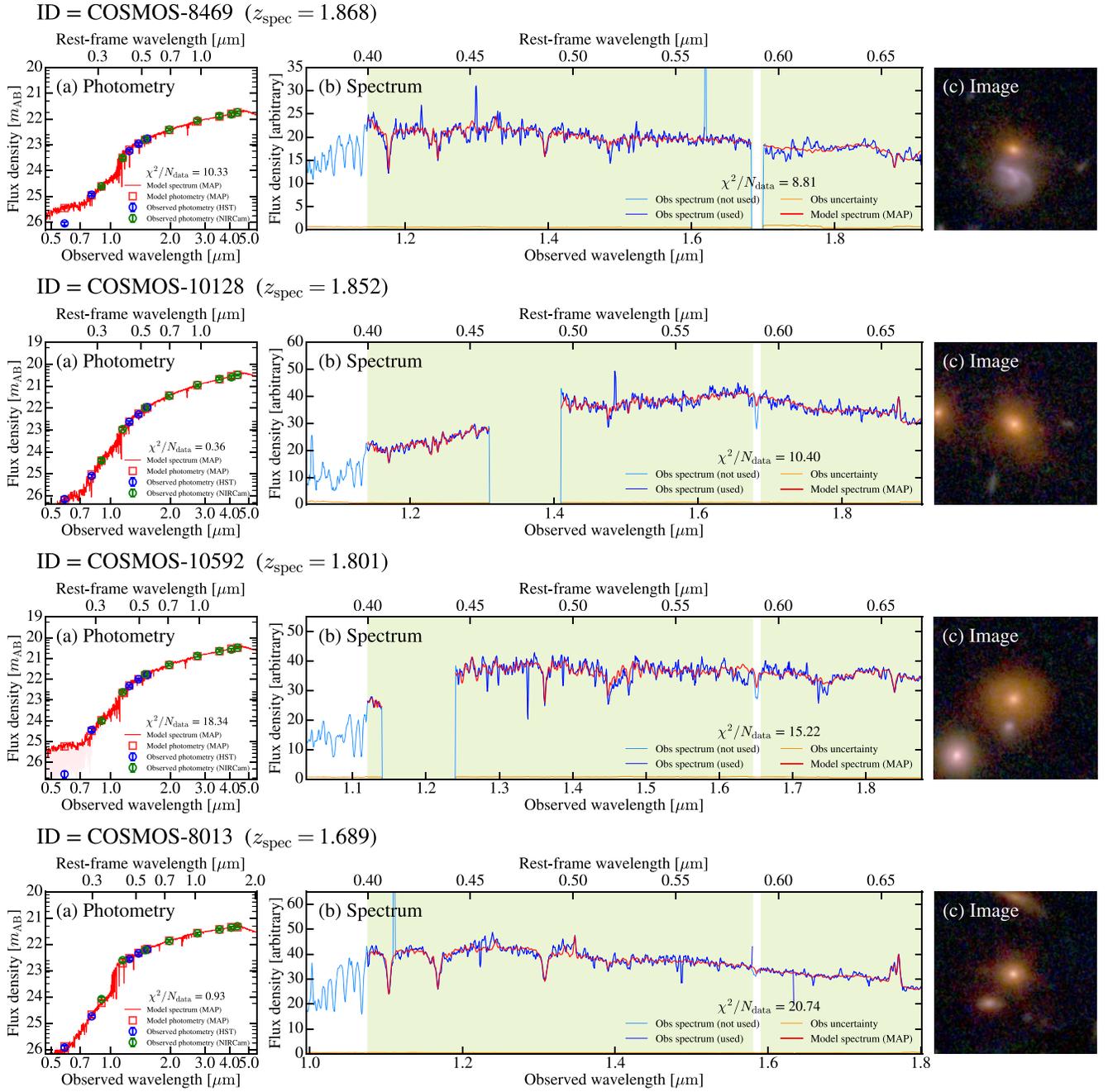

**Figure A4.** Continued.

## Appendix C
## SFH Inferred Using the Bursty Prior

In Section 3.2, we refit all our 14 quiescent galaxies using the bursty SFH prior and check how the inferred SFHs change with the assumed prior. Tables A1 and A2 summarize the results of Prospector fitting using the bursty prior. Figure A5 compares: (a) the formation redshift ($z_{\rm form}$); (b) the quenching redshift ($z_{\rm quench}$); and (c) the light-weighted ages of all 14 quiescent galaxies derived using the continuity and bursty priors. The error bars indicate the 16th–84th percentiles of the parameters from 1000 random posteriors. Broadly, the two epochs and the light-weighted ages are consistent. The overall larger error bars in the y-axis show that the SFHs are generally less constrained when the bursty prior is used.

**Table A1**
Prospector Fitting Results of 14 Massive Quiescent Galaxies in Our Sample Using the Bursty SFH Prior

| (1) 3D-HST COSMOS ID | (2) $z_{\rm spec}$ | (3) $\log(M_\star/M_\odot)$ | (4) [Fe/H] | (5) $\hat{\tau}_{\rm dust,old}$ | (6) $\hat{\tau}_{\rm dust,young}$ | (7) $n$ |
|---|---|---|---|---|---|---|
| 7549  | 2.626 | $10.81^{+0.06}_{-0.06}$ | $-0.68^{+0.15}_{-0.16}$ | $0.83^{+0.17}_{-0.14}$ | $0.79^{+0.27}_{-0.23}$ | $0.06^{+0.19}_{-0.18}$ |
| 21477 | 2.473 | $10.67^{+0.04}_{-0.04}$ | $-0.25^{+0.19}_{-0.24}$ | $0.74^{+0.10}_{-0.12}$ | $0.89^{+0.19}_{-0.17}$ | $0.25^{+0.09}_{-0.11}$ |
| 11142 | 2.444 | $10.91^{+0.04}_{-0.05}$ | $-0.03^{+0.07}_{-0.09}$ | $1.16^{+0.08}_{-0.07}$ | $1.14^{+0.33}_{-0.27}$ | $-0.26^{+0.07}_{-0.07}$ |
| 10565 | 2.442 | $10.84^{+0.05}_{-0.04}$ | $-0.24^{+0.07}_{-0.07}$ | $0.76^{+0.10}_{-0.10}$ | $0.86^{+0.22}_{-0.20}$ | $0.15^{+0.13}_{-0.16}$ |
| 10339 | 2.363 | $10.45^{+0.03}_{-0.03}$ | $-0.20^{+0.12}_{-0.15}$ | $0.65^{+0.09}_{-0.09}$ | $0.66^{+0.19}_{-0.19}$ | $-0.38^{+0.16}_{-0.17}$ |
| 9395  | 2.127 | $10.72^{+0.03}_{-0.03}$ | $-0.13^{+0.07}_{-0.07}$ | $0.35^{+0.07}_{-0.06}$ | $0.36^{+0.13}_{-0.11}$ | $-0.15^{+0.18}_{-0.29}$ |
| 10400 | 2.099 | $10.35^{+0.04}_{-0.06}$ | $-0.16^{+0.18}_{-0.15}$ | $0.09^{+0.09}_{-0.06}$ | $0.09^{+0.09}_{-0.06}$ | $-0.12^{+0.24}_{-0.24}$ |
| 11494 | 2.092 | $11.76^{+0.04}_{-0.05}$ | $0.08^{+0.06}_{-0.04}$ | $0.47^{+0.06}_{-0.06}$ | $0.44^{+0.15}_{-0.12}$ | $0.13^{+0.10}_{-0.15}$ |
| 18668 | 2.086 | $11.10^{+0.05}_{-0.07}$ | $-0.09^{+0.11}_{-0.10}$ | $1.38^{+0.14}_{-0.13}$ | $1.28^{+0.25}_{-0.25}$ | $0.09^{+0.20}_{-0.16}$ |
| 16419 | 1.925 | $11.37^{+0.06}_{-0.04}$ | $-0.07^{+0.05}_{-0.03}$ | $0.51^{+0.18}_{-0.07}$ | $0.46^{+0.14}_{-0.12}$ | $0.17^{+0.17}_{-0.16}$ |
| 8469  | 1.868 | $10.54^{+0.04}_{-0.04}$ | $-0.09^{+0.12}_{-0.12}$ | $0.52^{+0.09}_{-0.10}$ | $0.52^{+0.13}_{-0.11}$ | $0.12^{+0.13}_{-0.12}$ |
| 10128 | 1.852 | $11.23^{+0.05}_{-0.06}$ | $-0.50^{+0.12}_{-0.09}$ | $0.81^{+0.12}_{-0.12}$ | $0.78^{+0.26}_{-0.22}$ | $0.16^{+0.14}_{-0.16}$ |
| 10592 | 1.801 | $11.18^{+0.05}_{-0.05}$ | $0.13^{+0.07}_{-0.06}$ | $0.22^{+0.09}_{-0.08}$ | $0.22^{+0.13}_{-0.11}$ | $0.07^{+0.22}_{-0.22}$ |
| 8013  | 1.690 | $10.59^{+0.03}_{-0.04}$ | $-0.19^{+0.14}_{-0.14}$ | $0.67^{+0.11}_{-0.11}$ | $0.62^{+0.20}_{-0.18}$ | $-0.26^{+0.10}_{-0.09}$ |

**Note.** The table format is the same as in Table 1. Column (1): 3D-HST COSMOS ID. Column (2): fitted spectroscopic redshift. Column (3): stellar mass of a galaxy. Column (4): metallicity. Column (5): optical depth (in V band) for the diffuse dust component (see the details in C. Conroy et al. 2009). Column (6): optical depth for the birth-cloud component. Column (7): the power-law modifier to the shape of the D. Calzetti et al. (2000) dust attenuation curve (see the details in M. Kriek & C. Conroy 2013).

**Table A2**
Properties of 14 Massive Quiescent Galaxies, Derived Using the Bursty Prior, Classified into Three Groups by Their SFHs

| (1) ID | (2) 3D-HST COSMOS ID | (3) $z_{\rm obs}$ | (4) $\log(M_\star/M_\odot)$ | (5) $\mathrm{age}_{\rm light}$ (Gyr) | (6) $z_{\rm form}$ | (7) $t_{\rm form}^{\rm lbt}$ (Gyr) | (8) $z_{\rm quench}$ | (9) $t_{\rm quench}^{\rm lbt}$ (Gyr) | (10) $v_{\rm quench}$ (dex Gyr$^{-1}$) |
|---|---|---|---|---|---|---|---|---|---|
| Group-1: Early Formation and Early Quenching | | | | | | | | | |
| 1 | 7549  | 2.626 | 10.81 | $2.03^{+0.26}_{-0.35}$ | $14.75^{+10.52}_{-6.58}$ | $2.24^{+0.15}_{-0.35}$ | $9.70^{+6.29}_{-5.73}$ | $2.02^{+0.25}_{-1.08}$ | $4.01^{+6.17}_{-1.90}$ |
| 2 | 10400 | 2.099 | 10.35 | $2.33^{+0.48}_{-0.50}$ | $9.39^{+9.21}_{-4.36}$ | $2.65^{+0.32}_{-0.66}$ | $4.20^{+3.93}_{-1.47}$ | $1.70^{+0.84}_{-0.94}$ | $3.44^{+4.50}_{-1.52}$ |
| 3 | 16419 | 1.925 | 11.37 | $2.31^{+0.20}_{-0.14}$ | $19.99^{+0.80}_{-7.50}$ | $3.27^{+0.01}_{-0.17}$ | $12.78^{+0.00}_{-5.17}$ | $3.11^{+0.00}_{-0.35}$ | $3.03^{+1.27}_{-0.55}$ |
| 4 | 10128 | 1.825 | 11.23 | $2.17^{+0.30}_{-0.34}$ | $5.97^{+5.88}_{-1.14}$ | $2.63^{+0.57}_{-0.29}$ | $2.83^{+0.93}_{-0.94}$ | $1.26^{+0.64}_{-1.19}$ | $4.57^{+27.72}_{-2.43}$ |
| 5 | 10592 | 1.801 | 11.18 | $2.30^{+0.40}_{-0.38}$ | $8.26^{+8.90}_{-3.62}$ | $3.05^{+0.40}_{-0.68}$ | $2.35^{+9.82}_{-0.47}$ | $0.85^{+2.46}_{-0.69}$ | $4.00^{+10.40}_{-2.07}$ |
| Group-2: Early Formation and Recent Quenching | | | | | | | | | |
| 6  | 10565 | 2.442 | 10.84 | $1.03^{+0.13}_{-0.09}$ | $4.05^{+0.82}_{-0.21}$ | $1.18^{+0.31}_{-0.10}$ | $2.75^{+0.16}_{-0.17}$ | $0.32^{+0.14}_{-0.17}$ | $10.08^{+11.16}_{-4.43}$ |
| 7  | 9395  | 2.127 | 10.72 | $0.98^{+0.10}_{-0.11}$ | $3.40^{+0.13}_{-0.25}$ | $1.24^{+0.08}_{-0.17}$ | $2.38^{+0.00}_{-0.15}$ | $0.34^{+0.00}_{-0.19}$ | $13.24^{+10.80}_{-4.84}$ |
| 8  | 11494 | 2.092 | 11.76 | $1.99^{+0.18}_{-0.27}$ | $12.70^{+4.51}_{-5.85}$ | $2.84^{+0.12}_{-0.45}$ | $2.34^{+0.00}_{-0.09}$ | $0.34^{+0.00}_{-0.11}$ | $12.25^{+9.07}_{-5.16}$ |
| 9  | 18668 | 2.086 | 11.10 | $0.98^{+0.26}_{-0.21}$ | $3.37^{+1.19}_{-0.45}$ | $1.29^{+0.57}_{-0.34}$ | $2.20^{+0.28}_{-0.06}$ | $0.16^{+0.35}_{-0.09}$ | $16.97^{+19.95}_{-11.35}$ |
| 10 | 8469  | 1.868 | 10.54 | $0.85^{+0.16}_{-0.14}$ | $3.21^{+0.43}_{-0.16}$ | $1.53^{+0.27}_{-0.12}$ | $1.90^{+0.03}_{-0.01}$ | $0.05^{+0.06}_{-0.02}$ | $14.68^{+16.18}_{-7.88}$ |
| Group-3: Recent Formation and Recent Quenching | | | | | | | | | |
| 11 | 21477 | 2.473 | 10.67 | $0.38^{+0.13}_{-0.07}$ | $2.86^{+0.21}_{-0.03}$ | $0.39^{+0.17}_{-0.03}$ | $2.56^{+0.21}_{-0.03}$ | $0.10^{+0.21}_{-0.03}$ | $12.91^{+10.54}_{-4.56}$ |
| 12 | 11142 | 2.444 | 10.91 | $0.25^{+0.05}_{-0.05}$ | $2.72^{+0.10}_{-0.04}$ | $0.29^{+0.09}_{-0.04}$ | $2.53^{+0.05}_{-0.05}$ | $0.10^{+0.05}_{-0.05}$ | $16.81^{+17.33}_{-7.59}$ |
| 13 | 10339 | 2.363 | 10.45 | $0.59^{+0.10}_{-0.08}$ | $2.97^{+0.11}_{-0.06}$ | $0.61^{+0.09}_{-0.05}$ | $2.65^{+0.16}_{-0.16}$ | $0.32^{+0.15}_{-0.17}$ | $12.75^{+12.41}_{-5.35}$ |
| 14 | 8013  | 1.690 | 10.59 | $0.58^{+0.16}_{-0.16}$ | $1.98^{+0.13}_{-0.11}$ | $0.54^{+0.21}_{-0.18}$ | $1.77^{+0.04}_{-0.05}$ | $0.17^{+0.08}_{-0.10}$ | $13.63^{+13.25}_{-5.09}$ |

**Note.** Column (1): ID. Column (2): 3D-HST COSMOS ID. Column (3): observed redshift. Column (4): stellar mass. Column (5): light-weighted stellar age (in Gyr). Column (6): formation redshift. Column (7): formation epoch in lookback time (in Gyr). Column (8): quenching redshift. Column (9): quenching epoch in lookback time (in Gyr). Column (10): quenching speed (dex Gyr$^{-1}$). All the results are based on Prospector fitting using the bursty prior.

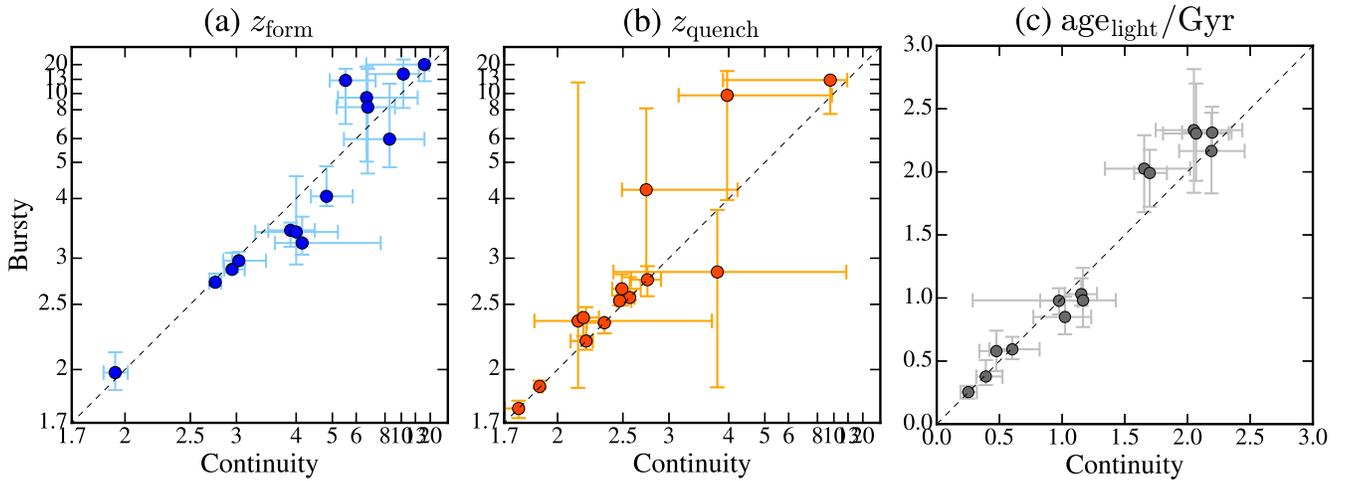

**Figure A5.** Comparison of (a) the formation redshift ($z_{form}$), (b) the quenching redshift ($z_{quench}$), and (c) the light-weighted ages (in Gyr) of the 14 massive quiescent galaxies derived using the continuity and bursty SFH priors. The error bars indicate the 16th to 84th percentiles of the derived parameters. We find that the formation and quenching epochs and the light-weighted ages inferred using the two priors are broadly consistent.

## Appendix D
## SFH Recovery Tests

In this section, we test whether the past starburst event can be recovered from SED fitting using the two SFH priors (the continuity and bursty priors). We generate mock photometry and a mock spectrum using FSPS by feeding a PSB-like SFH pushed back by 0.0, 0.5, 1.0, and 2.0 Gyr from the observed epochs to mimic what the data would look like when a PSB is observed a few Gyr after quenching. For the PSB-like SFH, the MAP SFH of COMSOS-11142 is used. The same best-fit stellar and dust parameters are used to generate the mock SED, and we did not consider dust evolution with time. For the mock photometry, we use the same HST/ACS+WFC3 and JWST/NIRCam filters we used for Blue Jay fitting and assume the same signal-to-noise ratio (S/N) of COSMOS-11142 for each filter. To provide the same fitting conditions as for the observed galaxies, we multiply the mock spectrum by the same polynomial to generate an uncalibrated mock spectrum and allow a polynomial to fit. To generate mock noise, we take the ratio between the observed spectrum and the best-fit model from COSMOS-11142 and multiply this ratio with the mock uncalibrated spectrum to generate a noise-added mock spectrum. The flux error is calculated assuming the same S/N at each pixel.

Figure A6 shows the mock data and the results of the Prospector fitting. We first fit both the mock photometry and spectrum using the continuity SFH prior. From top to the bottom, the figure shows the mock data using the SFH of a PSB pushed back by 0.0, 0.5, 1.0, and 2.0 Gyr from the observed epochs. The first column shows the mock photometry (blue: HST/ACS+WFC3; green: JWST/NIRCam filters) and the best-fit SED model in red. The second column shows the mock uncalibrated spectrum in blue and the mock errors in orange, and the red line shows the best-fit spectrum. The third column shows the input (blue line) and fitted SFHs (green shaded regions). The black dashed line is the MAP, and the green line shows the 50th percentile of the posterior distribution. The light-weighted ages of the stellar populations with the input and fitted SFHs are given in blue and green at the lower right corner of the rightmost panel.

The history of a past starburst and rapid quenching, even only a few hundred Myr before the observed epoch, appears to be difficult to reconstruct from the SED fitting using the continuity SFH prior. The derived light-weighted ages agree well with the ages of the stellar population with the input SFHs. The past starburst pushed by 0.5 Gyr is recovered but with a much lower intensity. When observed 1 Gyr after quenching, the reconstructed SFH looks different from the input SFH; it had roughly constant SFRs until recently and is rapidly quenched, just like the galaxies in Group-2. Finally, the past starburst history pushed by 2 Gyr is also not recovered at all. This might suggest that the galaxies in our Group-1 and Group-2 could instead be PSBs but observed a few Gyr after quenching.

We further investigate whether the inability to reproduce the history of a past starburst and rapid quenching is due to the continuity SFH prior. We refit the same mock data but this time use the bursty SFH prior. This prior does not necessarily favor abrupt changes but allows changes more flexibly. We do not show the best-fit SED with the bursty SFH prior, as it is very similar to the best fit from the continuity SFH prior (the red lines in Figure A6).

The past starbursts appear to be much better reproduced with the bursty SFH prior. The fourth column in Figure A6 shows the reconstructed SFHs from the mock data using the bursty SFH prior. In the same format as the third column, the figure shows the input (blue line) and fitted SFHs (pink shaded regions). The black dashed line is the MAP SFH, and the magenta line shows the 50th percentile of the posterior distribution. The history of a starburst and rapid quenching pushed back by 0.5, 1, and 2 Gyr is much better reproduced with the bursty prior. However, while the presence of the past bursts is better reproduced using the bursty prior, the exact moment and the intensity of the star formation peak is still not well recovered if it is more than 2 Gyr ago. The inferred SFH using the bursty prior when the PSB is observed 2 Gyr after quenching (the rightmost panel in the last row in Figure A6) shows that the reconstructed star formation peak is off by ~0.5 Gyr from the input star formation peak; the reconstructed stellar population is slightly younger. This suggests that the early formation (>2 Gyr) we infer from the reconstructed SFHs could have a systematic error up to ~0.5 Gyr.

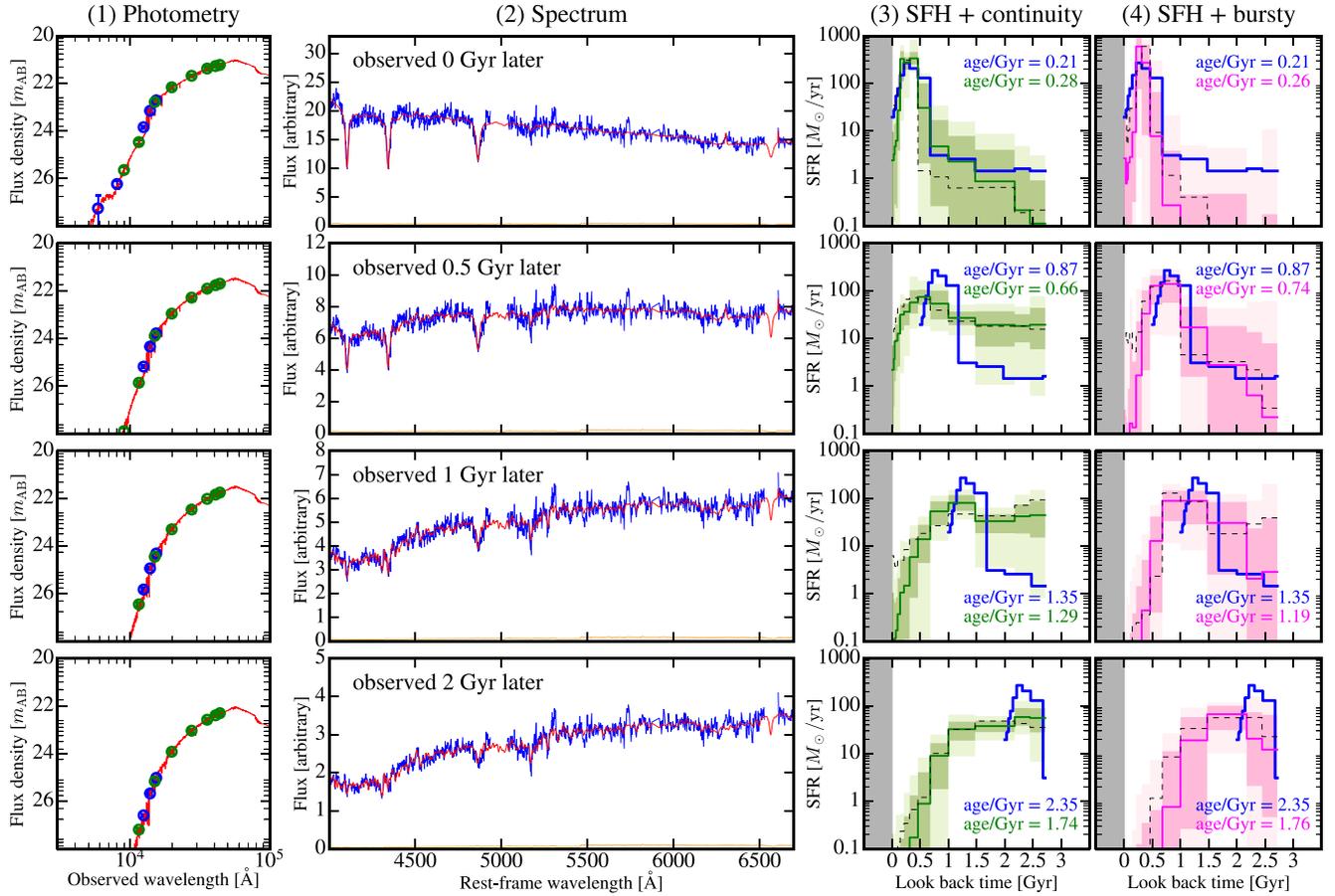

**Figure A6.** Test showing whether past starbursts can be recovered from SED fitting. We generate mock photometry and a mock spectrum using FSPS with a PSB-like SFH pushed by 0.0, 0.5, 1.0, and 2.0 Gyr, presented from the top to the bottom rows. The first column shows the mock photometry in circles. The same HST/ACS +WFC3 (blue circles) and JWST/NIRCam filters (green circles) are used for mock photometry. The best-fit SED model is shown with the red line. The second column shows the noise-added mock spectrum generated with each SFH, and the red line shows the best-fit SED model. The blue line in each panel in the third and fourth columns is the input SFH, and the light-weighted age of the stellar population is shown in the right corner. The reconstructed SFHs from Prospector fitting are shown in the third (using the continuity prior) and the fourth (using the bursty prior) columns. The stellar ages inferred from SED fitting are also given in the right corner. The green and yellow–green shades in the third column include $1\sigma$ and $2\sigma$ of the posteriors at each SFH time bin. The green line shows the 50th percentile of the posteriors at each time bin, and the dashed black line shows the SFH from MAP. The same format is used with a magenta color for the reconstructed SFHs using the bursty prior in the fourth column. We find that the continuity prior does not seem to well reproduce the past starburst events. The past starburst events are much better reproduced when the bursty prior is used. However, even the bursty prior cannot perfectly reproduce the exact moment and intensity of the past starbursts that happened more than 2 Gyr ago.

## Appendix E
## Prospector Fitting Results for the Two Oldest Galaxies

We present Prospector fitting results for the two oldest galaxies, COSMOS-7549 (Figure A7) and 16419 (Figure A8). The figures are in a similar format as in Figure 1 showing the observed photometry (panel a) and NIRSpec data (panel b). The best-fit spectrum is shown as a red line. The composite color image is shown in panel c, in a $5'' \times 5''$ box. Panel d shows nonparametric SFHs inferred using the continuity SFH prior (green, left) and the bursty SFH prior (magenta, right). Panel e shows the posterior distributions of the fitted stellar and dust parameters (green and pink histograms from the fitting using the continuity and the bursty prior, respectively).

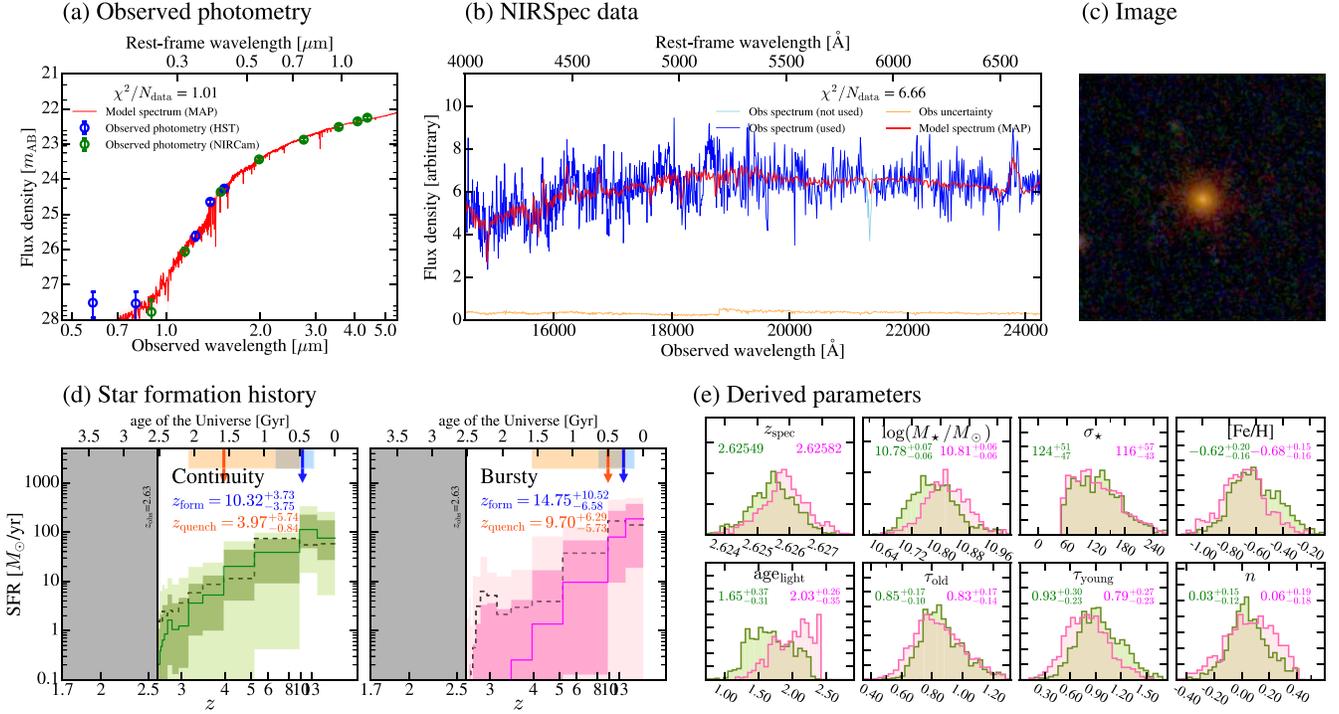

**Figure A7.** Prospector fitting results of COSMOS-7549, in a similar format as Figure 1. (a) Observed HST and NIRCam photometry shown as blue and green circles, respectively, and the red line shows the best-fit SED. (b) NIRSpec data are shown with a blue line, and the best-fit spectrum is again shown with a red line. The orange line shows the flux uncertainty. (c) Color-composite image of the galaxy in $5'' \times 5''$ box. (d) Nonparametric SFH derived from fitting using the continuity SFH prior (left) and the bursty SFH prior (right). Each SFH is the same as shown in Figures 4 and 5, respectively. (e) Posterior distribution of the fitted stellar and dust parameters; the green histogram is from the fitting using the continuity prior and the pink histogram is from the bursty prior.

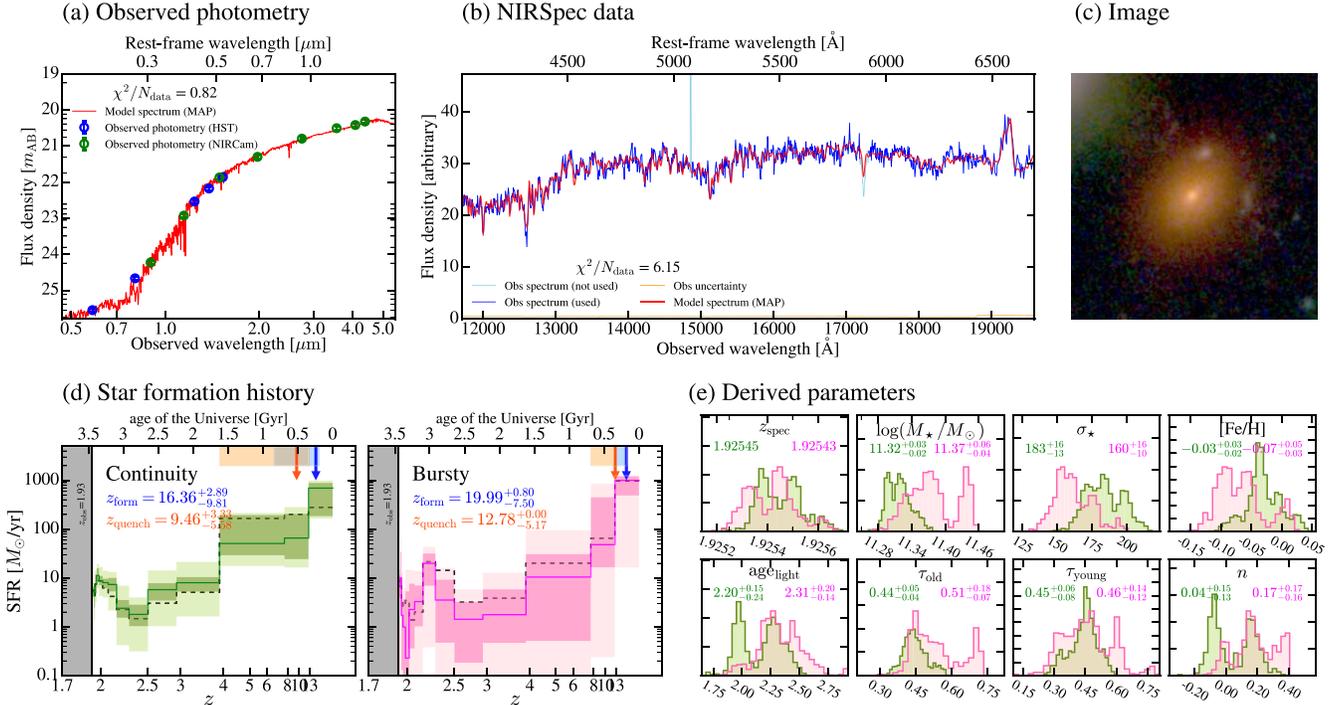

**Figure A8.** Prospector fitting results of 16419. The same format as Figure A7.

ORCID iDs

Minjung Park https://orcid.org/0000-0002-8435-9402
Sirio Belli https://orcid.org/0000-0002-5615-6018
Charlie Conroy https://orcid.org/0000-0002-1590-8551
Benjamin D. Johnson https://orcid.org/0000-0002-9280-7594
Rebecca L. Davies https://orcid.org/0000-0002-3324-4824
Joel Leja https://orcid.org/0000-0001-6755-1315
Sandro Tacchella https://orcid.org/0000-0002-8224-4505
J. Trevor Mendel https://orcid.org/0000-0002-6327-9147
Chloë Benton https://orcid.org/0000-0001-5378-9998
Razieh Emami https://orcid.org/0000-0002-2791-5011
Yijia Li https://orcid.org/0000-0002-0682-3310
Elijah P. Mathews https://orcid.org/0000-0003-0384-0681
Rohan P. Naidu https://orcid.org/0000-0003-3997-5705
Erica J. Nelson https://orcid.org/0000-0002-7524-374X
Bryan A. Terrazas https://orcid.org/0000-0001-5529-7305
Rainer Weinberger https://orcid.org/0000-0001-6260-9709

## References

Alberts, S., Williams, C. C., Helton, J. M., et al. 2024, ApJ, 975, 85
Antwi-Danso, J., Papovich, C., Esdaile, J., et al. 2023, arXiv:2307.09590
Baldry, I. K., Glazebrook, K., Brinkmann, J., et al. 2004, ApJ, 600, 681
Barnes, J. E., & Hernquist, L. E. 1991, ApJL, 370, L65
Bekki, K., Couch, W. J., Shioya, Y., & Vazdekis, A. 2005, MNRAS, 359, 949
Belli, S., Newman, A. B., & Ellis, R. S. 2019, ApJ, 874, 17
Belli, S., Park, M., Davies, R. L., et al. 2024, Natur, 630, 54
Bluck, A. F. L., Mendel, J. T., Ellison, S. L., et al. 2014, MNRAS, 441, 599
Bugiani, L., Belli, S., Park, M., et al. 2024, arXiv:2406.08547
Byler, N., Dalcanton, J. J., Conroy, C., & Johnson, B. D. 2017, ApJ, 840, 44
Calzetti, D., Armus, L., Bohlin, R. C., et al. 2000, ApJ, 533, 682
Cameron, E. 2011, PASA, 28, 128
Cargile, P. A., Conroy, C., Johnson, B. D., et al. 2020, ApJ, 900, 28
Carnall, A. C., McLure, R. J., Dunlop, J. S., et al. 2023a, Natur, 619, 716
Carnall, A. C., McLeod, D. J., McLure, R. J., et al. 2023b, MNRAS, 520, 3974
Carnall, A. C., Cullen, F., McLure, R. J., et al. 2024, MNRAS, 534, 325
Chabrier, G. 2005, Astrophysics and Space Science Library, 327 (Berlin: Springer), 41
Charlot, S., & Fall, S. M. 2000, ApJ, 539, 718
Choi, J., Dotter, A., Conroy, C., et al. 2016, ApJ, 823, 102
Concas, A., Popesso, P., Brusa, M., Mainieri, V., & Thomas, D. 2019, A&A, 622, A188
Conroy, C., & Gunn, J. E. 2010, ApJ, 712, 833
Conroy, C., Gunn, J. E., & White, M. 2009, ApJ, 699, 486
Conselice, C. J., Rajgor, S., & Myers, R. 2008, MNRAS, 386, 909
Croton, D. J., Springel, V., White, S. D. M., et al. 2006, MNRAS, 365, 11
Davies, R. L., Belli, S., Park, M., et al. 2024, MNRAS, 528, 4976
Davies, R. L., Förster Schreiber, N. M., Lutz, D., et al. 2020, ApJ, 894, 28
de Graaff, A., Setton, D. J., Brammer, G., et al. 2024, arXiv:2404.05683
Dekel, A., & Burkert, A. 2014, MNRAS, 438, 1870
D'Eugenio, C., Daddi, E., Gobat, R., et al. 2020, ApJL, 892, L2
D'Eugenio, F., Perez-Gonzalez, P., Maiolino, R., et al. 2024, NatAs
Di Matteo, T., Springel, V., & Hernquist, L. 2005, Natur, 433, 604
Dome, T., Tacchella, S., Fialkov, A., et al. 2024, MNRAS, 527, 2139
Draine, B. T., & Li, A. 2007, ApJ, 657, 810
Dressler, A., & Gunn, J. E. 1983, ApJ, 270, 7
Dressler, A., Smail, I., Poggianti, B. M., et al. 1999, ApJS, 122, 51
Ferland, G. J., Porter, R. L., van Hoof, P. A. M., et al. 2013, RMxAA, 49, 137
Forrest, B., Marsan, Z. C., Annunziatella, M., et al. 2020, ApJ, 903, 47
Förster Schreiber, N. M., Übler, H., Davies, R. L., et al. 2019, ApJ, 875, 21
French, K. D. 2021, PASP, 133, 072001
French, K. D., Yang, Y., Zabludoff, A., et al. 2015, ApJ, 801, 1
Gallazzi, A., Bell, E. F., Zibetti, S., Brinchmann, J., & Kelson, D. D. 2014, ApJ, 788, 72
Gehrels, N. 1986, ApJ, 303, 336
Glazebrook, K., Nanayakkara, T., Schreiber, C., et al. 2024, Natur, 628, 277
Glazebrook, K., Schreiber, C., Labbé, I., et al. 2017, Natur, 544, 71
Goto, T. 2005, MNRAS, 357, 937
Gould, K. M. L., Brammer, G., Valentino, F., et al. 2023, AJ, 165, 248
Grogin, N. A., Kocevski, D. D., Faber, S. M., et al. 2011, ApJS, 197, 35
Gunn, J. E., & Gott, J. R. I. 1972, ApJ, 176, 1
Harvey, T., Conselice, C., Adams, N. J., et al. 2024, arXiv:2403.03908
Hinshaw, G., Larson, D., Komatsu, E., et al. 2013, ApJS, 208, 19
Hopkins, P. F., Hernquist, L., Cox, T. J., et al. 2006, ApJS, 163, 1
Ito, K., Tanaka, M., Valentino, F., et al. 2023, ApJL, 945, L9
Johnson, B. D., Leja, J., Conroy, C., & Speagle, J. S. 2021, ApJS, 254, 22
Kakimoto, T., Tanaka, M., Onodera, M., et al. 2024, ApJ, 963, 49
Kauffmann, G., Heckman, T. M., Tremonti, C., et al. 2003, MNRAS, 346, 1055
Kere, D., Katz, N., Weinberg, D. H., & Dave, R. 2005, MNRAS, 363, 2
Kewley, L. J., Heisler, C. A., Dopita, M. A., & Lumsden, S. 2001, ApJS, 132, 37
Koekemoer, A. M., Faber, S. M., Ferguson, H. C., et al. 2011, ApJS, 197, 36
Kriek, M., & Conroy, C. 2013, ApJL, 775, L16
Kriek, M., Price, S. H., Conroy, C., et al. 2019, ApJL, 880, L31
Lamperti, I., Harrison, C. M., Mainieri, V., et al. 2021, A&A, 654, A90
Leja, J., Carnall, A. C., Johnson, B. D., Conroy, C., & Speagle, J. S. 2019a, ApJ, 876, 3
Leja, J., Johnson, B. D., Conroy, C., et al. 2019b, ApJ, 877, 140
Leja, J., Speagle, J. S., Ting, Y.-S., et al. 2022, ApJ, 936, 165
Leung, G. C. K., Coil, A. L., Aird, J., et al. 2019, ApJ, 886, 11
Long, A. S., Antwi-Danso, J., Lambrides, E. L., et al. 2024, ApJ, 970, 68
Looser, T. J., D'Eugenio, F., Maiolino, R., et al. 2023, arXiv:2306.02470
Looser, T. J., D'Eugenio, F., Maiolino, R., et al. 2024, Natur, 629, 53
Lovell, C. C., Roper, W., Vijayan, A. P., et al. 2023, MNRAS, 525, 5520
Madau, P., & Dickinson, M. 2014, ARA&A, 52, 415
Man, A., & Belli, S. 2018, NatAs, 2, 695
McConachie, I., Wilson, G., Forrest, B., et al. 2022, ApJ, 926, 37
Mihos, J. C., & Hernquist, L. 1996, ApJ, 464, 641
Muzzin, A., Marchesini, D., Stefanon, M., et al. 2013, ApJ, 777, 18
Nanayakkara, T., Glazebrook, K., Jacobs, C., et al. 2024, NatSR, 14, 3724
Narayanan, D., Lower, S., Torrey, P., et al. 2024, ApJ, 961, 73
Pacifici, C., Kassin, S. A., Weiner, B. J., et al. 2016, ApJ, 832, 79
Park, M., Belli, S., Conroy, C., et al. 2023, ApJ, 953, 119
Peng, Y., Maiolino, R., & Cochrane, R. 2015, Natur, 521, 192
Rodriguez-Gomez, V., Pillepich, A., Sales, L. V., et al. 2016, MNRAS, 458, 2371
Roy, N., Bundy, K., Rubin, K. H. R., et al. 2021, ApJ, 919, 145
Schreiber, C., Glazebrook, K., Nanayakkara, T., et al. 2018, A&A, 618, A85
Setton, D. J., Dey, B., Khullar, G., et al. 2023, ApJL, 947, L31
Setton, D. J., Khullar, G., Miller, T. B., et al. 2024, ApJ, 974, 145
Sherman, S., Jogee, S., Florez, J., et al. 2020, MNRAS, 499, 4239
Silk, J., & Rees, M. J. 1998, A&A, 331, L1
Skelton, R. E., Whitaker, K. E., Momcheva, I. G., et al. 2014, ApJS, 214, 24
Slob, M., Kriek, M., Beverage, A. G., et al. 2024, ApJ, 973, 131
Snyder, G. F., Cox, T. J., Hayward, C. C., Hernquist, L., & Jonsson, P. 2011, ApJ, 741, 77
Somerville, R. S., Hopkins, P. F., Cox, T. J., Robertson, B. E., & Hernquist, L. 2008, MNRAS, 391, 481
Steinhardt, C. L., Jespersen, C. K., & Linzer, N. B. 2021, ApJ, 923, 8
Suess, K. A., Kriek, M., Price, S. H., & Barro, G. 2020, ApJL, 899, L26
Suess, K. A., Leja, J., Johnson, B. D., et al. 2022, ApJ, 935, 146
Suess, K. A., Williams, C. C., Robertson, B., et al. 2023, ApJL, 956, L42
Sun, Y., Lee, G.-H., Zabludoff, A. I., et al. 2024, MNRAS, 528, 5783
Tacchella, S., Conroy, C., Faber, S. M., et al. 2022a, ApJ, 926, 134
Tacchella, S., Dekel, A., Carollo, C. M., et al. 2016, MNRAS, 457, 2790
Tacchella, S., Diemer, B., Hernquist, L., et al. 2019, MNRAS, 487, 5416
Tacchella, S., Finkelstein, S. L., Bagley, M., et al. 2022b, ApJ, 927, 170
Tacchella, S., Johnson, B. D., Robertson, B. E., et al. 2023, MNRAS, 522, 6236
Tanaka, M., Onodera, M., Shimakawa, R., et al. 2024, ApJ, 970, 59
Trussler, J., Maiolino, R., Maraston, C., et al. 2020, MNRAS, 491, 5406
Valentino, F., Brammer, G., Gould, K. M. L., et al. 2023, ApJ, 947, 20
Valentino, F., Tanaka, M., Davidzon, I., et al. 2020, ApJ, 889, 93
Veilleux, S., Maiolino, R., Bolatto, A. D., & Aalto, S. 2020, A&ARv, 28, 2
Wang, B., Leja, J., Atek, H., et al. 2024, ApJ, 963, 74
Wang, B., Leja, J., Bezanson, R., et al. 2023, ApJL, 944, L58
Weinberger, R., Springel, V., Pakmor, R., et al. 2018, MNRAS, 479, 4056
Whitaker, K. E., Franx, M., Leja, J., et al. 2014, ApJ, 795, 104
Whitaker, K. E., Kriek, M., van Dokkum, P. G., et al. 2012a, ApJ, 745, 179

Whitaker, K. E., van Dokkum, P. G., Brammer, G., & Franx, M. 2012b, ApJL, 754, L29
White, S. D. M., & Rees, M. J. 1978, MNRAS, 183, 341
Wild, V., Almaini, O., Dunlop, J., et al. 2016, MNRAS, 463, 832
Wild, V., Taj Aldeen, L., Carnall, A., et al. 2020, MNRAS, 494, 529
Wu, P.-F., van der Wel, A., Bezanson, R., et al. 2018, ApJ, 868, 37
Wuyts, S., Förster Schreiber, N. M., van der Wel, A., et al. 2011, ApJ, 742, 96
Yuan, T., Nanayakkara, T., Kacprzak, G. G., et al. 2014, ApJL, 795, L20
Zabludoff, A. I., Zaritsky, D., Lin, H., et al. 1996, ApJ, 466, 104
Zheng, Y., Dave, R., Wild, V., & Montero, F. R. 2022, MNRAS, 513, 27
Zolotov, A., Dekel, A., Mandelker, N., et al. 2015, MNRAS, 450, 2327